\documentclass[superscriptaddress,amsmath,amssymb,aps,prl,twocolumn,floatfix]{revtex4-2}

\usepackage{graphicx}
\usepackage{bm}
\usepackage[colorlinks, linkcolor=mycol, citecolor=mycol, urlcolor=mycol, breaklinks]{hyperref}
\usepackage{braket}
\usepackage{bbold}
\usepackage{xcolor} 
\usepackage{comment}
\usepackage{physics}
\usepackage{verbatim}

\newcommand{\mi}{ {\rm i} }
\newcommand{\me}{ {\rm e} }

\usepackage{amsmath}
\usepackage{amssymb}
\definecolor{mycol}{RGB}{10,55,130}
\definecolor{annoyingpink}{RGB}{255,0,211}

\begin{document}
\title{Rise and fall, and slow rise again, of operator entanglement under dephasing} 

\author{D.~Wellnitz}
\affiliation{ISIS (UMR 7006) and CESQ, CNRS and Universit\'{e} de Strasbourg, 67000 Strasbourg, France}
\affiliation{IPCMS (UMR 7504), CNRS, 67000 Strasbourg, France}

\author{G.~Preisser}
\affiliation{ISIS (UMR 7006) and CESQ, CNRS and Universit\'{e} de Strasbourg, 67000 Strasbourg, France}

\author{V.~Alba}
\affiliation{Dipartimento di Fisica, Universit\`a di Pisa, and INFN Sezione di Pisa,
Largo Bruno Pontecorvo 3, Pisa, Italy}

\author{J.~Dubail}
\affiliation{Universit\'{e} de Lorraine, CNRS, LPCT, F-54000 Nancy, France}
\affiliation{ISIS (UMR 7006) and CESQ, CNRS and Universit\'{e} de Strasbourg, 67000 Strasbourg, France}

\author{J.~Schachenmayer}
\thanks{schachenmayer@unistra.fr}
\affiliation{ISIS (UMR 7006) and CESQ, CNRS and Universit\'{e} de Strasbourg, 67000 Strasbourg, France}
\affiliation{IPCMS (UMR 7504), CNRS, 67000 Strasbourg, France}

\date{\today}

\begin{abstract}
The operator space entanglement entropy, or simply `operator entanglement' (OE), is an indicator of the complexity of quantum operators and of their approximability by Matrix Product Operators (MPO). We study the OE of the density matrix of 1D many-body models undergoing dissipative evolution. It is expected that, after an initial linear growth reminiscent of unitary quench dynamics, the OE should be suppressed by dissipative processes as the system evolves to a simple stationary state. Surprisingly, we find that this scenario breaks down for one of the most fundamental dissipative mechanisms: dephasing. Under dephasing, after the initial `rise and fall' the OE can rise again, increasing logarithmically at long times. Using a combination of MPO simulations for chains of infinite length and analytical arguments valid for strong dephasing, we demonstrate that this growth is inherent to a $U(1)$ conservation law. We argue that in an XXZ spin-model and a Bose-Hubbard model the OE grows universally as $\frac{1}{4} \log_2 t$ at long times, and as $\frac{1}{2} \log_2 t$ for a Fermi-Hubbard model. We trace this behavior back to anomalous classical diffusion processes.
\end{abstract}

\maketitle

The study of quantum many-body systems through the prism of their quantum entanglement continues to prove extremely fruitful~\cite{amico2008entanglement,eisert2010colloquium}. In particular, the growth of entanglement in time-evolving quantum many-body systems is of fundamental interest~\cite{calabrese2005evolution,fagotti2008evolution,znidaric2008many,alba2017entanglement,jonay2018coarse,lukin2019probing}: not only is it useful to characterize the dynamics, but the amount of entanglement also indicates whether a quantum evolution can be efficiently simulated on a classical computer. In one dimension (1D) the connection can be made via the concept of matrix product states (MPS) \cite{vidal2004efficient,verstraete2008matrix,schollwock2011density-matrix,paeckel2019time-evolution}. An MPS is a  decomposition of a many-body state vector into a product of $\chi \times \chi$ matrices (where the entries of the matrices are local kets). In such a representation, the bipartite von Neumann entanglement entropy $S$ is bounded by $\max[S] = \log_2(\chi)$. Consequently, to represent a physical state $\left| \psi(t) \right>$ with entanglement entropy $S(t)$ as an MPS, the matrix size (or `bond dimension') has to grow at least as $\chi \propto 2^{S(t)}$ with time. For example, an evolution where $S$ increases linear in time can therefore be considered computationally hard~\cite{schuch2008entropy}. 

The past few years have seen the arrival of novel experiments capable of synthetically engineering quantum many-body models in controllable and clean environments, e.g.~using optically trapped ultracold atoms~\cite{bloch2012quant,adams2019rydberg,browaeys2020many,morgado2021quantum}, molecules~\cite{gadway2016strong}, or ions~\cite{blatt2012quant}. Since such experiments are currently bringing the goal of analog quantum simulation into sight~\cite{cirac2012goals,georgescu2014quantum}, the question of entanglement growth, and thus classical simulability, has become very important. 

\begin{figure}
	\centering
    \includegraphics[width=\linewidth]{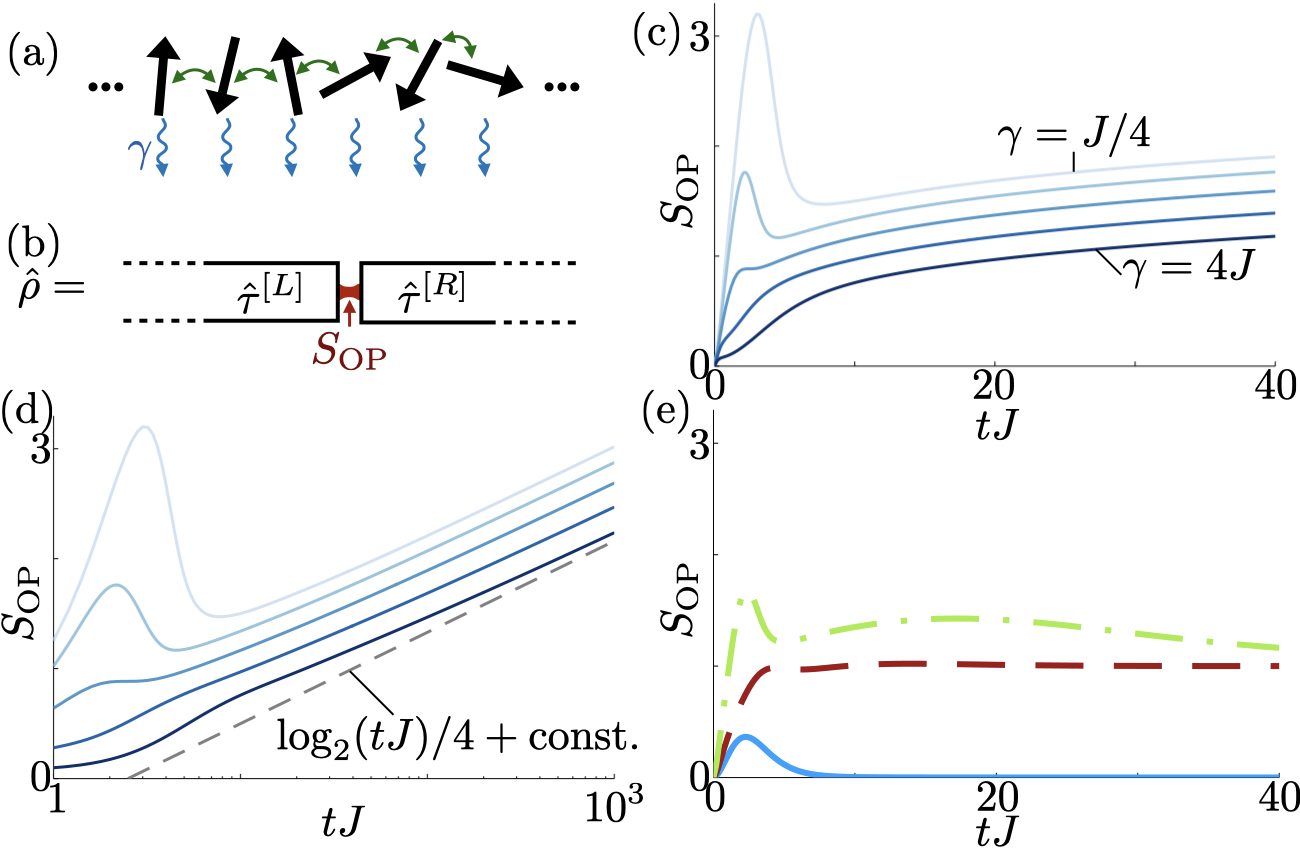}
	\caption{(a) We compute dynamics of spin chains with coherent nearest-neighbor couplings (double-arrows) and local dephasing at rate $\gamma$ (wiggly arrows). (b) We analyze the growth of OE ($S_\mathrm{OP}$) for a bipartition of an infinite chain into a left/right block [from a Schmidt decomposition of the density matrix $\hat \rho$, see Eq.~\eqref{eq:bipart}]. (c) Time evolution of $S_\mathrm{OP}$ for a N\'eel state in the XXZ model for different values of $\gamma = J/4,J/2,J,2J,4J$ in order of increasing darkness (anisotropy $J_z=-J$). (d) Same as (c), demonstrating logarithmic growth at long times (log-scale time axis). Grey dashed: Analytic long-time prediction: $S_\mathrm{OP} = \log_2(Jt)/4 + \mathrm{const}$. (e) $S_\mathrm{OP}$ for dynamics in models breaking magnetization conservation. Green dash-dotted: XYZ model ($J_x = J, J_y = 0.8J, J_z = -J/2, \gamma=J/2$), Red dashed: transverse field Ising model ($h_z = J, \gamma=J/2$), blue solid: XXZ model with initial N\'eel state in $x$-direction ($J_z=-J/2, \gamma=J/2$). Results converged for time step $\Delta t J = 0.2$ [$\Delta t J = 0.5$ for panel (d) at long times] and different values for the bond dimension $\chi=256,512,1024$ (see Appendix for convergence).}
	\label{fig:fig1}
\end{figure}

Every experiment has small couplings to its environment, and should therefore be considered as an open quantum system described by a density matrix $\hat \rho$. Analogously to MPSs for pure states, also a matrix product operator (MPO) form of the density matrix $\hat{\rho}$ can be defined. An MPO form allows to easily express the density matrix as Schmidt decomposition between a left and a right block:
\begin{align}
    \hat \rho = \sum_{a} \lambda_{a} \hat \tau^{[L]}_a \hat \tau^{[R]}_a,
    \label{eq:bipart}
\end{align}
with $\mathrm{Tr}(\hat \tau^{[L/R]}_a \hat \tau^{[L/R]}_b) = \delta_{ab}$, and $\lambda_a$  the Schmidt coefficients [schematically this is depicted in Fig.~\ref{fig:fig1}(b)]. The bipartite entropy of this decomposition is given by the `operator space entanglement entropy' or simply `operator entanglement' (OE) defined as~\cite{zanardi2000entangling,zanardi2001entanglement,wang2002quantum,prosen2007operator, dubail2017entanglement,zhou2017operator,jonay2018coarse, alba2019operator,wang2019barrier,styliaris2021information}
\begin{align} \label{eq:sop}
    S_{\rm OP} = - \sum_{a} \lambda_{a}^2 \log_2\qty(\lambda_{a}^2).
\end{align}
The OE quantifies how many Schmidt values are at least needed for faithfully approximating decomposition~\eqref{eq:bipart}, and thus indicates the efficiency of an MPO representation \footnote{Note that here this indicates approximability w.r.t.~the two-norm of the vectorized density matrix.}. It can be easily computed numerically~\cite{prosen2007operator,zhou2017operator,alba2019operator,noh2020efficient,rakovszky2020dissipation} and is amenable to analytical treatment~\cite{dubail2017entanglement,bertini2020operatorI,bertini2020operatorII}. We stress that OE is not necessarily connected to genuine quantum entanglement between distinct blocks of spins when $\hat \rho$ is a mixed state. Still, it is a crucial quantity as it puts severe restrictions on the possibility to approximate $\hat{\rho}$ by an MPO. Furthermore, OE can give insights into quantum many body effects such as quantum chaos and information scrambling~\cite{zhou2017operator,wang2019barrier,styliaris2021information}.

\smallskip

Here, we analyze the far-from-equilibrium dynamics of the OE, $S_\mathrm{OP}$, in open many-body quantum systems. Our models include coherent nearest-neighbor Hamiltonian couplings that compete with incoherent single-particle dephasing~\cite{rossini2021coherent,cai2013algebraic,medvedyeva2016exact,foss2017solvable,znidaric2015relaxation} at rate $\gamma$ [see sketch in Fig.~\ref{fig:fig1}(a)].
Under dephasing, fluctuations and coherences can decay towards equilibrium in a universal algebraic and sub-diffusive manor~\cite{poletti2012interaction,poletti2013emergence,ren2020noise,bouganne2020anomalous}.
We treat dissipation in a Lindblad master equation. Dephasing arises due to a coupling with the environment in which local magnetization is preserved, e.g.~for laser driven transitions due to laser-phase fluctuations~\cite{plankensteiner2016laser,gardiner2004quantum}, or due to spontaneous photo-absorption of lattice photons in optical lattices~\cite{pichler2010nonequilibrium}.
We compute the evolution of $S_\mathrm{OP}$ for an infinite MPO representation of the full density matrix using an infinite time evolving block decimation algorithm (iTEBD) with re-orthogonalization~\cite{orus2008infinite}.

\smallskip

Surprisingly, we find that for the magnetization conserving XXZ model and well-defined initial magnetization [see text below Eq.~\eqref{eq:ham} for the definition of our models], the OE exhibits a universal logarithmic growth at long times [see~Fig.~\ref{fig:fig1}(c/d)]. An identical universal behavior is also observed for particle number conserving Bose- and Fermi-Hubbard models [see discussion below]. Strikingly, as shown in Fig.~\ref{fig:fig1}(e), this logarithmic growth breaks down if the symmetry is broken by initial states, or for Liouvillians without magnetization conservation (see Appendix for more examples). In the latter scenarios $S_\mathrm{OP}$ saturates, or even vanishes at long times. In this paper we explain this behavior by considering symmetry-resolved OE in combination with known results for classical models of interacting particles~\cite{harris1965diffusion,levitt1973dynamics,arratia1983motion}.

\medskip

\paragraph{Model ---} We focus on infinite spin-1/2 chains, which evolve under the general Hamiltonian ($\hbar \equiv 1$):
\begin{equation}
\hat H \!=\! \frac{1}{4} \sum^{}_{i}  \left( J_x \hat \sigma^x_i \hat \sigma^x_{i+1} \!+\! J_y \hat \sigma^y_{i} \hat \sigma^y_{i+1} \!+\! J_z \hat \sigma^z_i \hat \sigma^z_{i+1}\right) + \frac{h_z}{2} \sum_i \hat \sigma^z_i.
\label{eq:ham}
\vspace{0.5em}
\end{equation}
Here, $\hat \sigma^{x,y,z}_i$ denote standard Pauli matrices defined in a local basis $\ket{\downarrow,\uparrow}_i$, $J_{x,y,z}$ are the respective nearest-neighbor spin couplings, and $h_z$ is a field strength along the $z$ direction. Our Hamiltonian~\eqref{eq:ham} includes: i) the \textit{XXZ model}, with $J_x=J_y\equiv J$, $h_z=0$; ii) an \textit{XYZ model}, with $J_x\equiv J$, $J_x\neq J_y$, $h_z=0$; and iii) a \textit{transverse Ising model},  with $J_x\equiv J$, $J_y=J_z=0$, $h_z \neq 0$. We are interested in the dynamics of highly-excited states. Here, we choose pure N\'eel product states polarized along the $z$ direction, $\hat \rho_0 = \ket{\psi_0}\bra{\psi_0}$ with $\ket{\psi_0} = \bigotimes_i \ket{\uparrow}_{2i-1}\ket{\downarrow}_{2i}$, or a tilted N\'eel state along the $x$ direction, $\ket{\psi_0}= \bigotimes_i\ket{\rightarrow}_{2i-1}\ket{\leftarrow}_{2i}$ with $\ket{\leftrightarrows}_i = (\ket{\uparrow}_i \mp \ket{\downarrow}_i)/\sqrt{2}$. Dynamics  is governed by a Lindblad master equation,
 \begin{align} \label{eq:master}
\frac{d}{dt} \hat \rho = -\mi [\hat H, \hat \rho] + \sum_i \mathcal{D}^{[i]} \hat \rho  \equiv  \mathcal L \hat \rho \, ,
\end{align}
with the local dephasing super-operators
$ \mathcal{D}^{[i]} \hat \rho = \gamma/2 (\hat \sigma_i^z \hat  \rho \hat \sigma_i^z - \hat \rho)$ and the Liouvillian super-operator $\mathcal L$.

\medskip

\paragraph{MPO decomposition and OE ---} For $L$ spins, the full many-body density matrix $\hat \rho$ of a spin-$1/2$ system is a $2^L \times 2^L$ hermitian matrix with unit trace. The amount of information encoded in $\hat\rho$ can be effectively compressed using matrix product decompositions~\cite{schollwock2011density-matrix,paeckel2019time-evolution}. This can be done in different ways~\cite{weimer2021simulation}: For instance, by decomposing $\hat \rho$ into a particular (not unique) statistical mixture of pure states, while using an MPS for the latter. Then, the Lindblad  dynamics can  be computed using quantum trajectories~\cite{daley2014quantum}. Alternatively, one can use a direct MPO representation for $\hat\rho$, e.g.~simply by effectively vectorizing local density matrices~\cite{zwolak2004mixed}, or by constructing MPOs in a locally purified form which preserves positivity~\cite{verstraete2004matrix,werner2016positive}.

\begin{figure*}
	\centering
    \includegraphics[width=\linewidth]{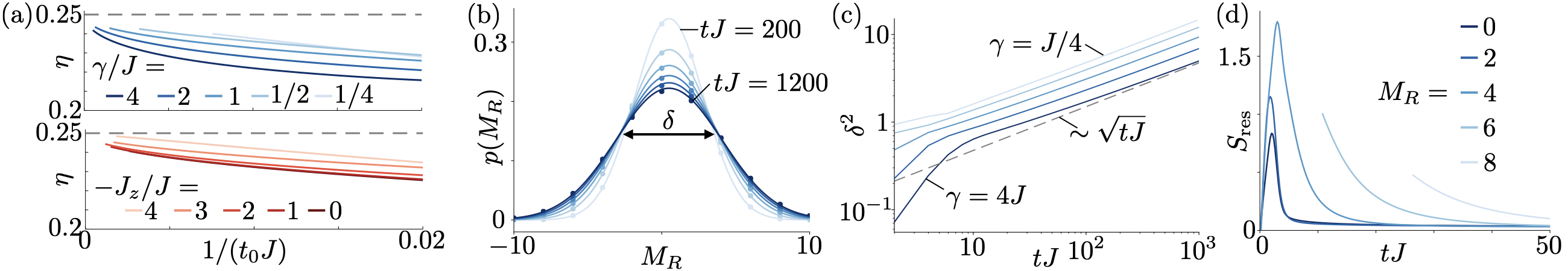}
	\caption{(a) Numerical determination of $\eta$ for long times. $\eta$ at time $t_0 J$ is obtained as the local tangent to $S_\mathrm{OP}(t) = S_0 + \eta \log(tJ)$ at $t_0 J$. We find $\eta \to 1/4$ (grey dashed line) for all parameters and $t_0\to \infty$. Top (blue): Fixed $J_z=-J$ and various $\gamma/J=1/4,1/2,1,2,4$. Bottom (red): Fixed $\gamma=J$ and various $-J_z/J=0,1,2,3,4$.
	(b) Probabilities $p(M_R)$ for right-half magnetization $M_R$ of the infinite chain (see text) at increasingly late times ($200 \leq tJ \leq 1200$ from light to dark, $J_z=-J$, $\gamma=J/2$. Lines are Gaussian fits.
	(c) Variance $\delta^2$ of Gaussian fits [as in (b)] as a function of time for $J_z=-J$ and $\gamma/J=1/4,1/2,1,2,4$ (light to dark). The grey dashed line indicates $\delta^2 \sim \sqrt{tJ}$ (double-log scale).
	(d) Symmetry-resolved operator entanglement entropies $S_\mathrm{res}$ as a function of time (see text). For short times and the larger $M_R=6,8$ the probabilities $p(M_R)$ are exponentially suppressed and no sub-machine-precision data could be extracted. Same parameters as panel (b).
	Results converged for $\Delta t J = 0.5$ [(a)-(c)] and $\Delta t J = 0.1$ [(d)] and different values of $\chi=512,1024,2048$~(see Appendix).
	}
	\label{fig:fig2}
\end{figure*}

\smallskip

Here, we decompose $\hat \rho$ into a canonical MPO form~\cite{zwolak2004mixed}, which is formally achieved by an iterative application of the Schmidt decomposition from Eq.~\eqref{eq:bipart}, until each spin $n$ is described by a matrix of unique local operators $\hat \gamma_{a_n,a_{n+1}}^{[n]}$, i.e.~$\hat \rho = \sum_{\{a_n\}} \prod_n \lambda_{a_n} \hat \gamma_{a_n,a_{n+1}}^{[n]}$. By choosing local basis operators, $\hat e_{i_n}$, for the density matrix of spin $n$, we can then decompose:
\begin{align}
\label{eq:mpo-G}
    \hat \rho = \sum_{\{i_n\}} \sum_{\{a_n\}}^\chi \prod_n \Gamma_{a_n a_{n+1}}^{[n]\,i_n} \lambda_{a_{n}}^{[n]} \bigotimes_n \hat e_{i_n}, 
\end{align} 
where $\Gamma^{[n]}$ are three-dimensional tensors and  $\lambda^{[n]}$ the Schmidt vectors. Tensors are truncated at a maximum MPO bond dimension $\chi$. All results shown are numerically converged in $\chi$~(see Appendix). For $\hat e_i$ we choose local eigenoperators for the multiplication with $\hat \sigma^z$ from the left and right, to take advantage of the magnetization conservation (see below). Both the initial state and the Hamiltonian are invariant under translation by two lattice sites. As a consequence, in Eq.~\eqref{eq:mpo-G} one has $\Gamma^{[n+2]} = \Gamma^{[n]}$ and $\lambda^{[n+2]} = \lambda^{[n]}$~\cite{vidal2007classical,orus2008infinite}, and only two $\Gamma$ and $\lambda$ tensors are needed to encode a density matrix. The time evolution is then computed with a fourth order Trotter decomposition of the matrix exponential of the Liouvillian $\exp(\mathcal L \Delta t)$~\cite{zwolak2004mixed,orus2008infinite,Sornborger_Higher_1999}.
Importantly, since the dynamics is non-unitary, a na\"ive implementation of this algorithm destroys the orthogonality of the decomposition in Eq.~\eqref{eq:mpo-G}, such that with time, the $\lambda$s do not correspond to orthogonal Schmidt bases anymore. We fix this by re-orthogonalizing the tensors after updates~\cite{orus2008infinite}.

\smallskip

When considering the XXZ model, the total magnetization $\hat S^z = \sum_n \hat \sigma_n^z$ is conserved. This means that $\hat \rho$ stays an eigenoperator of $\hat S^z$ in the sense that $\hat S^z \hat \rho(t) = M \hat \rho(t)$ at all times (for the N\'eel state, $M=0$). Note that alternatively, one could also define a condition for multiplication from the right. Due to magnetization preservation, the $\hat\tau_a^{\scriptscriptstyle[R]}$ matrices in Eq.~\eqref{eq:bipart} can be chosen to be eigenoperators of the `right-half magnetization' of the chain, $\hat S^z_R\equiv \sum_{n>0} \hat \sigma_n^z$ (w.l.o.g.~we define the right half as $n>0$), $\hat S^z_R\hat\tau_a^{\scriptscriptstyle[R]}= M_R \hat \tau_a^{\scriptscriptstyle[R]}$. Similarly, one can choose $\hat\tau_a^{\scriptscriptstyle[L]}$ to be eigenoperators of $\hat S^z_L \equiv \sum_{n\leq 0} \hat \sigma_n^z$ with $M_L=-M_R$. This means that the index $a$ in~Eq.~\eqref{eq:bipart} becomes a composite index $a\to (M_R,a')$, where $a'$ distinguishes the Schmidt coefficients corresponding to the same  $M_R$:
\begin{align}
\hat\rho=\sum_{M_R} \sqrt{p_{M_R}} \sum_{a'} \widetilde\lambda_{M_R, a'} \hat \tau^{[L]}_{-M_R,a'} \hat \tau^{[R]}_{M_R,a'} \,. \label{eq:bipartresolved}
\end{align}
Here we defined $\widetilde\lambda_{M_R,a}\equiv\lambda_{M_R,a}/\sqrt{p_{M_R}}$, with $p_{M_R}=\sum_{a}\lambda^2_{M_R,a}$ the probability of having magnetization $M_R$ in the right half. The existence of the conservation law makes our simulations much more efficient, since the block-diagonal form of the tensors can be exploited.

\medskip

\paragraph{Logarithmic increase of OE: Numerical results ---} In simulations in Fig.~\ref{fig:fig1} we noticed a distinctive different behavior of OE growth at long times (log-growth) for the magnetization conserving XXZ model compared to other models breaking this conservation law. Quite generically, at times  $t\ll\gamma^{-1}$, the dynamics is dominated by the Hamiltonian part in Eq.~(\ref{eq:master}). Sufficiently pure states at such short times can be approximated by the state  $\ket{\psi_t}=\me^{-\mi \hat H t}\ket{\psi_0}$. In that case the OE is simply twice the entanglement entropy of $|\psi_t\rangle$ (see e.g.~\cite{dubail2017entanglement}), and it is well established that the latter generically grows linearly in time (in the absence of disorder). At times $t \gtrsim \gamma^{-1}$, the initial coherence is destroyed by dephasing, and the OE decreases (see~Appendix for a more detailed discussion on the parameter dependence of the peak-heights). This `rise and fall' is clearly visible in Fig.~\ref{fig:fig1}, and it is typical for OE dynamics, and also other quantities such as the mutual information~\cite{carollo2021emergent,alba2021hydrodynamics}. Typically, under the dynamics in Eq.~\eqref{eq:master} the system is expected to relax to a simple stationary state characterized by the conserved quantities of Eq.~\eqref{eq:master}, or to the identity if there is no conservation law. The OE at late times converges towards the OE of that stationary state. This is visible in our simulations of the XYZ and Ising models, see Fig.~\ref{fig:fig1}(e). In this case only the parity $\hat{\Pi} =\bigotimes_i  \hat{\sigma}^z_i$ is preserved by the dynamics. Since the initial N\'eel state is an eigenstate of $\hat{\Pi}$, the stationary density matrix is a projector on a fixed parity sector, $\frac{1}{2} (1\pm\hat{\Pi})$, with the $\mathcal{O}(1)$ entropy $S_\mathrm{OP}=1 ~(=  \log_{\rm 2} 2)$. For the XXZ chain with the initial tilted N\'eel along the $x$ direction, even parity conservation is broken, and the stationary density matrix becomes the identity, $S_\mathrm{OP}=0$. In stark contrast, for the XXZ chain and initial N\'eel state, after the rise and fall dynamics, the OE increases again at long times, see Fig.~\ref{fig:fig1}(c), and this second increase is logarithmic in time, see Fig.~\ref{fig:fig1}(d). More precisely, we find the long-time behavior $S_{\mathrm{OP}}(t\to \infty) = \eta \log_2(tJ) + S_0$, which we will also understand analytically below. The prefactor $\eta$ converges universally to $\eta=1/4$ independent on the precise values of $\gamma$ and $J_z$, as shown in Fig.~\ref{fig:fig2}(a), and has also been observed with additional disorder~\cite{medvedyeva2016influence}. The offset $S_0$ depends on the characteristic time-scale of the long-time diffusive dynamics set by $J$, $J_z$ and $\gamma$~(see Appendix). Note that we find the evolution of OE in the XXZ model to be independent of the signs of $J$ and $J_z$.

\medskip

\begin{figure}
	\centering
    \includegraphics[width=\columnwidth]{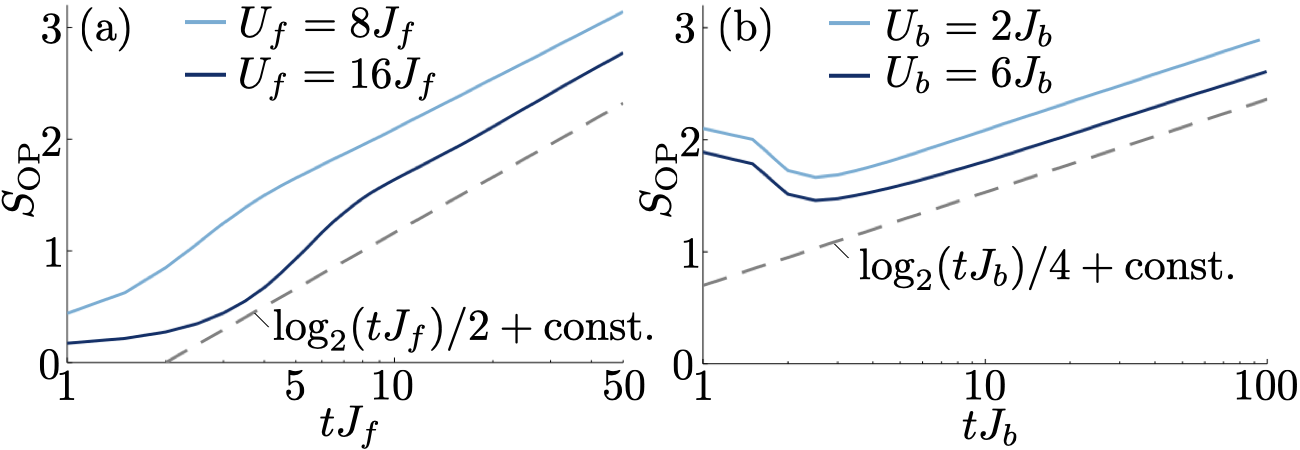}
	\caption{Logarithmic OE growth in the Fermi-Hubbard (a) and Bose-Hubbard (b) models with dephasing for different interaction strengths $U_{f/b}$. The grey dashed lines indicate the analytically expected long-time growth~(see Appendix). The initial states are: One fermion per site with alternating spins $\ket{\cdots\uparrow\downarrow\uparrow\cdots}$ (a) and alternating sites with $0$ or $1$ boson $\ket{\cdots 101 \cdots}$ (b). Parameters: $\gamma_f = 8J_f$ (a) and $\gamma_b = 2J_b$ (b), $\chi = 512,256$, $\Delta tJ_{f/b} = 1/2$, maximum bosons per site $n_\mathrm{max} = 4$.}
	\label{fig:fig3}
\end{figure}

\paragraph{Mechanism for logarithmic growth: Abelian symmetry and anomalous charge diffusion ---} To also analytically understand this behavior, we now consider the XXZ model evolution in 
the strong dephasing limit $\gamma \gg J$. The dissipators in the master equation~\eqref{eq:master} project the density matrix onto its diagonal part $\hat{\rho}_{\rm diag} = \sum_{\bm{\sigma}} \rho_{\bm{\sigma \sigma}} \ket{\bm{\sigma}}\bra{\bm{\sigma}}$, where the $\bm{\sigma}$ denote all binary vectors of spin-$z$ configurations. The dynamics then reduces to a classical master equation for the probability $p_{\bm{\sigma}} = \rho_{\bm{\sigma \sigma}}$, $d p_{\bm{\sigma}}/dt = \sum_{\bm{\sigma}'} \mathcal{M}_{\bm{\sigma \sigma}'} p_{\bm{\sigma}'}$. The stochastic matrix $\mathcal{M}$ was determined in Ref.~\onlinecite{cai2013algebraic} in second-order perturbation theory starting from Eq.~(\ref{eq:master}). It takes the form of an effective ferromagnetic Heisenberg Hamiltonian, $\mathcal M = -J^2/(8\gamma) \sum_i \left[ \hat{\sigma}^x_i \hat{\sigma}^x_{i+1}+\hat{\sigma}^y_i \hat{\sigma}^y_{i+1}+\hat{\sigma}^z_i \hat{\sigma}^z_{i+1}-1 \right]$. Importantly, $\mathcal{M}$ is the stochastic matrix of the Symmetric Simple Exclusion Process (SEP)~\cite{mallick2015exclusion,bernard2018transport}, a model of classical hard-core particles that perform random walks.

\smallskip

Crucially, in the SEP, the mean squared displacement of a tagged particle grows as $\left< X_t^2 \right> \propto \sqrt{t}$~\cite{levitt1973dynamics,arratia1983motion,lin2005random,derrida2009current,imamura2017large,grabsch2021closing}, as opposed to $\propto t$ for a usual random walk. This anomalous diffusion is universally found in problems of so-called `single-file diffusion'~\cite{hahn1996single,wei2000single,lin2005random}, when classical particles diffuse in one-dimensional channels without bypassing each other. Here, it is now also tied to anomalous scaling of the particle number fluctuations between the left and right half-systems. If $\Delta N(t) =M_R(t)/2$ is the excess number of particles (w.r.t.~the initial N\'eel state)  in the right half-system at time $t$, and if we tag the particle initially at the origin, then one can estimate $\Delta N(t) \simeq \rho_0 X_t$, where $\rho_0 = 1/2$ is the particle density in the N\'eel state. Consequently, $\left< \Delta N(t)^2 \right> \simeq \rho_0^2 \left< X_t^2\right>\propto \sqrt{t}$. More generally, the probability distribution of $M_R(t)$ is found to obey a scaling form at long times~\cite{derrida2009current}: $p(M_R(t)=m) \, \underset{t \rightarrow \infty}{\sim}  \exp \left( \sqrt{t}~ G \left( \frac{m}{\sqrt{t}} \right) \right)$. Here, the large-deviation function $G$ is non-positive, symmetric [$G(u) = G(-u)$], diverges when $|u| \rightarrow \infty$, and has a single minimum at $u=0$ (see~\cite{derrida2009current}). In particular, away from the tails the distribution is Gaussian with standard deviation $\delta = t^{1/4}/\sqrt{|G''(0)|}$. The Shannon entropy associated to number fluctuations is then
\begin{align}
    \label{eq:Snum}
 \nonumber     &S_{\rm num}(t) =  \sum_{m \in 2 \mathbb{Z}} -p(m) \log_2 p(m)  \\
\nonumber    &\quad \simeq   \int  -\sqrt{\frac{2}{ \pi \delta^2}} \me^{- \frac{ m^2}{2 \delta^2}} \log_2 \left( \sqrt{\frac{2}{ \pi \delta^2}} \me^{- \frac{ m^2}{2 \delta^2}} \right) \frac{dm}{2}  \\
    &\quad = \log_2 \delta + \log_2\qty(\sqrt{\pi \me/2} ) \, \underset{t \rightarrow \infty}{=} \, \frac{1}{4} \log_2  t  + \mathcal{O}(1) .
\end{align}
It is no coincidence that $S_{\rm OP}$ grows in the same way as the number fluctuations $S_{\rm num}$ at long times (see below).

\smallskip

Away from the $\gamma/J\gg 1$ limit, the XXZ chain no longer reduces to the SEP. Nevertheless we find that the same type of anomalous scaling persists. This is confirmed in Fig.~\ref{fig:fig2}, where we show that, for times accessible numerically, $p(M_R=m)$ is approximately Gaussian [Fig.~\ref{fig:fig2}(b)] with width $\delta \propto t^{\frac{1}{4}}$ [Fig.~\ref{fig:fig2}(c)]. Thus, even though the exact correspondence with the SEP breaks down at finite $\gamma/J$, the scaling of $S_{\rm num}$ in Eq.~\eqref{eq:Snum} remains unchanged. This result is also consistent with previous studies of transport in the XXZ and related models~\cite{znidaric2010dephasing,znidaric2010exact,eisler2011crossover,denardis2021subdiffusive}.

\medskip

\paragraph{Symmetry-resolved OE ---}We now show how the relation between $S_{\rm OP}$ and $S_{\rm num}$ can be revealed in so-called symmetry resolved operator entanglement. From Eq.~\eqref{eq:bipartresolved} we can derive a decomposition of the OE into the form:
\begin{align}
    S_{\rm OP} = \sum_{M_R} p_{M_R} S_{\rm res}(M_R) + S_{\rm num}(p_{M_R}) ,
    \label{eq:OEresolved}
\end{align}
where the `symmetry-resolved entanglement entropies' are $S_{\rm res}(M_R) = - \sum_a \widetilde\lambda^2_{M_R,a} \log_2( \widetilde\lambda^2_{M_R,a})$, and 
$S_{\rm num}$ is given in Eq.~\eqref{eq:Snum}. Such symmetry-resolved entropies have attracted attention recently~\cite{lukin2019probing,goldstein2018symmetry,xavier2018equipartition,parez2021quasiparticle,barghathi2018renyi,barghathi2019operationally}. In Fig.~\ref{fig:fig2}(d) we display $S_{\rm res}(M_R)$ for different values of $M_R$. Also $S_{\rm res}$ exhibits the `rise and fall' phenomenon discussed above, but independent of $M_R$ they decrease to very small values at late times. This means that the logarithmic increase $S_{\rm OP}$ is solely due to the growth of $S_{\rm num}$.

\medskip

\paragraph{Fermi- and Bose-Hubbard model ---} To demonstrate the generality of the logarithmic OE growth, we discuss two additional paradigmatic many-body setups featuring number conservation: i) a Fermi-Hubbard (FH) model, $\hat H_{\rm FH} = -J_f \sum_{n,\sigma} (\hat c_{\sigma, n}^\dagger \hat c_{\sigma, n+1} + \text{h.c.}) + U_f\sum_n \hat c_{\uparrow, n}^\dagger \hat c_{\downarrow, n}^\dagger \hat c_{\uparrow, n} \hat c_{\downarrow, n}$ with creation operators for spin-full fermions on site $n$, $\hat c_{\sigma, n}^\dagger$ ($\sigma = \uparrow, \downarrow$); and ii) a Bose-Hubbard (BH) model, $\hat H_{\rm BH} = -J_b \sum_n (\hat b_n^\dagger \hat b_{n+1} + \text{h.c.}) + U_b/2 \sum_n \hat b_n^\dagger \hat b_n^\dagger \hat b_n \hat b_n$, for bosons created by $\hat b_n^\dagger$.  In both cases we consider dephasing $\mathcal{D}^{[k]}\hat \rho = \gamma \hat L_k \rho \hat L_k^\dagger   - \gamma ( \hat L_k^\dag  \hat L_k \hat \rho +  \hat \rho  \hat L_k^\dag  \hat L_k)/2$, where $\hat L_{\sigma,n} = \hat c^\dag_{\sigma, n} \hat c_{\sigma, n}$ and $\hat L_n = \hat b^\dag_{n} \hat b_{n}$ in the FH and BH case, respectively. As demonstrated in Fig.~\ref{fig:fig3}, both models also exhibit a long-time logarithmic OE growth. In the FH model, we observe $S_\mathrm{OP} \sim \log_2(tJ_f)/2$. For $\gamma \gg J_f, U_f$, this can be understood analytically by considering the FH chain as the sum of two chains, one for each spin degree of freedom, both of which are described by the SEP and exhibit $S_\mathrm{OP} \sim \log_2(tJ_f)/4$. Here, interactions contribute only at higher orders~(see Appendix). The BH model exhibits $S_\mathrm{OP} \sim \log_2(tJ_b)/4$ analogous to the XXZ model. For large $U_b \gg \gamma \gg J_b$ the creation of doublons is energetically suppressed, leading back to the SEP. For finite interaction strength $(\gamma \gg J_b,U_b)$, a different classical limit is reached, which also features logarithmic OE growth with a prefactor close to $1/4$ stemming from a `Symmetric Inclusion
Process'~(see Appendix and Ref.~\cite{bernard2018transport}).

\medskip

\paragraph{Conclusion ---} We showed that in a dissipative system possessing a $U(1)$ conservation 
law the operator entanglement grows logarithmically at long times. We pinpointed the mechanism that leads to this logarithmic growth, and identified its prefactor with the [possibly anomalous] exponent characterizing the fluctuations of the charge associated with the $U(1)$ symmetry. Our results and methods are of general interest to studies of imperfect quantum computation and quantum simulation platforms, currently pushing into a regime where they may offer a quantum advantage. The entanglement entropy dynamics we study here connects directly to similar results obtained for discrete quantum circuit models~\cite{zhou2017operator,li2018quantum,chan2019unitary,noh2020efficient}, or to other dissipative simulation methods such as quantum trajectories~\cite{coppola2021growth,botzung2021engineered}. An understanding of the destructive processes of the environment on dynamics are essential, and the interplay between dissipation and coherent couplings can lead to interesting physics or state engineering~(e.g.~\cite{mark2020interplay,zhu2014suppressing,shchesnovich2010control,muller2022measurement}). In the future, it will be interesting to investigate how the presence of more complex symmetries such as SU(N) impacts entanglement dynamics.

\medskip

\paragraph{Acknowledgments} We thank P.~Calabrese, O.~Castro-Alvaredo, A.~Grabsch, G.~Misguich, S.~Murciano, G.~Pupillo and G. Sch\"utz for helpful discussions. This work was supported by LabEx NIE under contract ANR-11-LABX0058 NIE, and the QUSTEC program, which has received funding from the European Union's Horizon 2020 research and innovation program under the Marie Skłodowska-Curie grant agreement number 847471. This work is part of the Interdisciplinary Thematic Institute QMat, as part of the ITI 2021-2028 program of the
University of Strasbourg, CNRS and Inserm, and was supported by IdEx Unistra (ANR-10-IDEX-0002), SFRI STRAT'US project (ANR-20-SFRI-0012), and EUR QMAT ANR-17-EURE-0024 under the framework of the French
Investments for the Future Program. Our MPO codes make use of the intelligent tensor library (ITensor)~\cite{itensor}. Computations were carried  out using resources of the High Performance Computing Center of the University of Strasbourg, funded by Equip@Meso (as part of the Investments for the Future Program) and CPER Alsacalcul/Big Data.

\appendix

\setcounter{equation}{0}
\setcounter{figure}{0}
\setcounter{table}{0}
\setcounter{secnumdepth}{2}
\makeatletter
\renewcommand{\theequation}{S\arabic{equation}}
\renewcommand{\thefigure}{S\arabic{figure}}

\newpage

\begin{widetext}

\section*{Appendices}

In Sec.~\ref{sec:section_method_mpo}, we provide details on the numerical convergence. In Sec.~\ref{sec:noconservation}, we discuss a breaking of magnetization conservation on different levels. In Sec.~\ref{sec:peakheight} we show numerical results on the parameter dependence of the short time peaks of the OE. In Sec.~\ref{sec:oeoffset}, we discuss the perturbative estimate for the offset of the operator entanglement. In Sec.~\ref{sec:strongdephasing}, we identify the strong dephasing limits of the Fermi-Hubbard model, and of the Bose-Hubbard model, which are instrumental for explaining the prefactors of the logarithmic growth of the OE observed in the main text.

\section{Details on numerical convergence}
\label{sec:section_method_mpo}

\begin{figure}[b]
    \centering
    \includegraphics[width=1\textwidth]{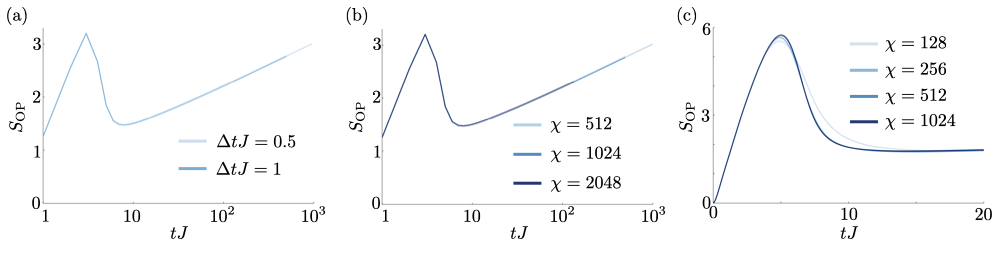}
    \caption{(a) Convergence in the time-step $\Delta t$ (in units $1/J$) for $\gamma=J/4$, $J_z = -J$, $\chi=1024$. (b) Convergence in $\chi$ for $\gamma=J/4$, $J_z = -J$, $\Delta t J = 0.5$. (c) Convergence in $\chi$ for $\gamma=J/8$, $J_z = -J$, $\Delta t J = 0.1$}
    \label{fig:conv}
\end{figure}

For solving the time-evolution with the iTEBD algorithm~\cite{vidal2007classical,orus2008infinite} we make use of a Trotter decomposition of the matrix exponential of the super-operator governing the full dissipative dynamics~\cite{zwolak2004mixed}. In order to reach long time-scales, we implement 4-th order decomposition~\cite{Sornborger_Higher_1999}, which allowed for time-step converged results even up to step sizes of $\Delta t = 1/(2J)$ [see Fig.~\ref{fig:conv}(a)]. $\Delta t$ signifies the full time step, which is composed of individual gates with time-steps of length $\Delta t/12$ and $\Delta t/6$. In particular, we use the method
\begin{align}
    (1)^T(1)(1)^T(-2)(1)^T(1)^T(1)^T(1)^T(1)(1)^T(1)(1)(1)(1)(-2)^T(1)(1)^T(1) \, .
\end{align}

\medskip

Throughout the paper we only show simulations up to times until which the results are converged in the MPO bond dimension $\chi$, defined in the truncated MPO decomposition in Eq.~(5). In practice, we ran simulations repeatedly, doubling bond dimensions in the different runs until lines become visually indistinguishable. We varied the bond dimension in the range $\chi=128,256,512,1024,2048$. Note that e.g.~for $\chi=1024$, a maximum possible entropy of $S_\mathrm{OP}\leq\log_2(\chi)=10$ is theoretically supported. Naturally, the evolution producing the largest OE required the largest values of $\chi$. The convergence plots in Fig.~\ref{fig:conv}(b) and Fig.~\ref{fig:conv}(c) demonstrate this procedure for ``worst-case'' scenarios (i.e.~the data with the largest OE values and for long times). We choose the data from Fig.~1(d) with $\gamma=J/4$, simulated up to long times, and the data for Fig.~3(a) with $J_z/J=-1$ and $\gamma=J/8$, which reached the largest values in the short time peak $S_\mathrm{OP} \sim 6$.

\medskip

In particular, the following parameters have been used for the figures in the main text:

\begin{center}
    \begin{tabular}{c||c|c|c|c}
        Figure & $-J_z/J$ & $\gamma/J$ & $\chi$ & $\Delta t J$ \\
        \hline
        Fig.~1(c) & 1 & $0.25$ & 512 & $0.2$ \\
        Fig.~1(c) & 1 & $\geq0.5$ & 256 & $0.1/0.2$ \\
        Fig.~1(d)(short times) & 1 & $\leq 1$ & 1024 & $0.1/0.2$ \\
        Fig.~1(d)(short times) & 1 & $> 1$ & 256 & $0.1/0.2$ \\
        Fig.~1(d)(long times) & 1 & all & 512 & $0.5$ \\
        Fig.~1(e) & all & all & 256 & $0.2$ \\
        Fig.~2(a)/S3(b) & 1 & $0.25$ & 2048 & $0.5$ \\
        Fig.~2(a)/S3(b) & 1 & $0.5$, $1$ & 1024 & $0.5$ \\
        Fig.~2(a)/S3(b) & 1 & 2, 4 & 512 & $0.5$ \\
        Fig.~2(a)/S3(b) & all & 1 & 1024 & $0.5$ \\
        Fig.~2(b) & 1 & 0.5 & 512 & $0.5$ \\
        Fig.~2(c) & 1 & all & 512 & $0.5$ \\
        Fig.~2(d) & 1 & 0.5 & 1024 & $0.1$ \\
        Fig.~3(a) & all & --- & 512 & 0.5 \\
        Fig.~3(b) & all & --- & 256 & 0.5 \\
        Fig.~S3(a) & all & $0.125$ & 1024 & $0.1$\\
        Fig.~S3(a) & all & $\geq0.25$ & 512 & $0.1/0.2$
    \end{tabular}
\end{center}
Whenever two $\Delta t$ values are given, different $\Delta t$s were used for different parameters for historic reasons, but  are clearly converged in either case.

\section{Breaking magnetization conservation}
\label{sec:noconservation}

\begin{figure}
    \centering
    \includegraphics[width=0.65\textwidth]{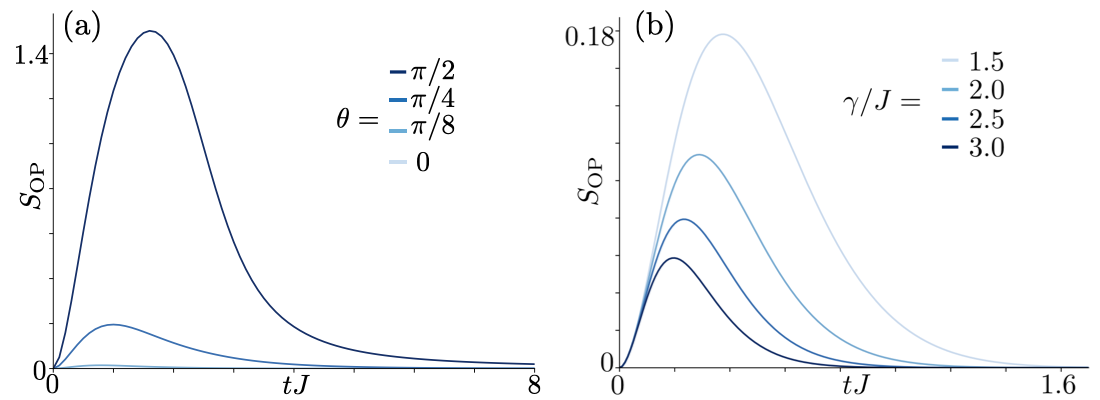}
    \caption{Rise and fall of operator entanglement without magnetization conservation. (a) We consider the XXZ model with dephasing, as in the main text, for $J_z = -J$, $\gamma = J/2$. The initial state is a fully polarized state $\bigotimes_n\mathrm{exp}[\mi \theta \hat \sigma_n^{(x)}]\ket{\downarrow}_n$, which are not and eigenstate of $\hat S_z$ (except for $\theta=0$). (b) We consider the XXZ model with $J_z = -J$ for a N\'eel initial state as in the main text. Instead of dephasing, we consider non-conserving dissipation $\mathcal{D}^\pm_n \hat \rho = \gamma/2 (2\hat \sigma_n^\pm \hat \rho \sigma_n^\mp -  \hat \sigma_n^\mp \hat \sigma_n^\pm \hat \rho -   \hat \rho \hat \sigma_n^\mp \hat \sigma_n^\pm)$, and for various rates $\gamma$. Simulations for $\chi = 128$, $\Delta t J = 0.1$}
    \label{fig:noconservation}
\end{figure}

In this section, we give a few more examples without magnetization conservation. First, we discuss fully polarized states that are tilted at an angle $\theta$, $\bigotimes_n\mathrm{exp}[\mi \theta \hat \sigma_n^{(x)}]\ket{\downarrow}_n$. For $0 < \theta < \pi$, these states are not eigenstates of $\hat S_z$. We find that after the initial growth, the OE decays to zero, just like for the initial tilted N\'eel state. For the special cases $\theta = 0(\pi)$, the state $\ket{\cdots \downarrow\downarrow\downarrow \cdots}$ ($\ket{\cdots\uparrow\uparrow\uparrow\cdots}$) is an eigenstate of the Liouvillian, and thus the OE remains zero at all times.

If magnetization conservation is broken on the level of the Liouvillian, we find a similar rise and fall of OE, which remains zero at long times [see Fig.~\ref{fig:noconservation}(b)]. For the case of both an incoherent pump and and decay at rate $\gamma$, after the initial rise and fall, we expect the steady state to be the (trivial) infinite temperature state with $S_\mathrm{OP} = 0$.

\section{Short-time peak heights}
\label{sec:peakheight}

\begin{figure}[h]
	\centering
    \includegraphics[width=0.7\columnwidth]{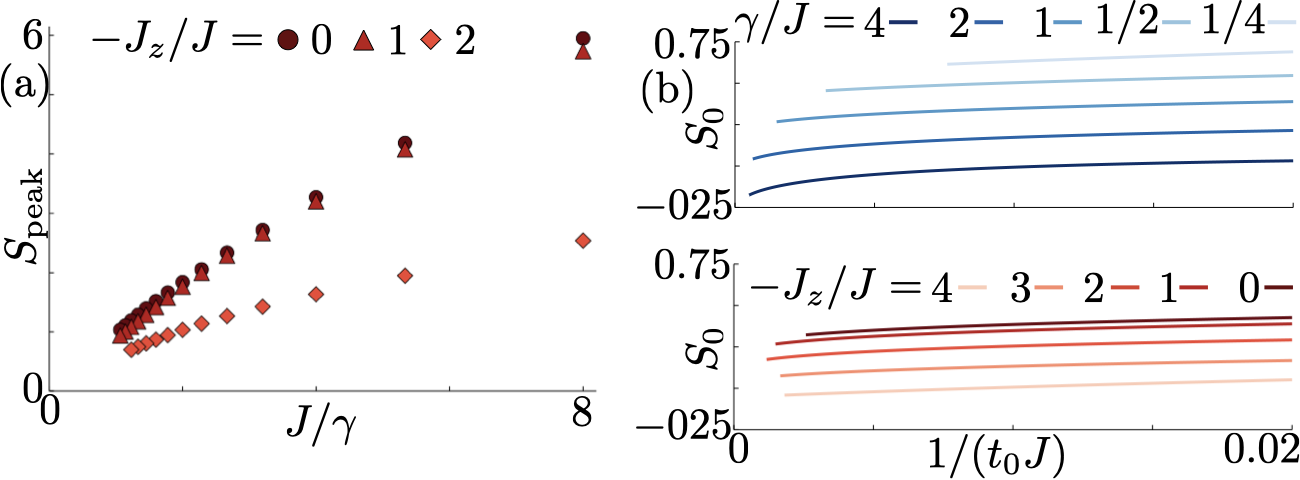}
	\caption{(a) Short-time peak-height of $S_\mathrm{OP}$ as a function of $1/\gamma$ for different $J_z$.
	(b) Offset $S_0$ (extracted from fits as in Fig.~2 in the main manuscript) as a function of the inverse tangent time $1 / (t_0J)$. Blue lines (top) indicate variable $\gamma$ at fixed $J_z = -J$, red lines (bottom) indicate variable $J_z$ at fixed $\gamma$. Results converged for (a) $\Delta t J = 0.1$ and (b) $\Delta t J = 0.5$, and different values of $\chi=512,1024,2048$}
	\label{fig:peakheights}
\end{figure}

\medskip

Whether a state can be efficiently approximated by an MPO is determined by the OE of that state, which has a local maximum at short times. Thus, until not too long times, the approximability of the dynamics is determined by the value at this maximum $S_\mathrm{peak}$. This value is shown in Fig.~\ref{fig:peakheights}(a) as a function of the dephasing rate $\gamma$ and the interaction strength $J_z$. We find that generally the peak-height grows as $S_{\rm peak}\sim 1/\gamma$ for small $\gamma$. Interestingly, $S_{\rm peak}$ decreases when increasing the spin interaction strength $|J_z|$. As a consequence, the dynamics under strong dephasing can always be simulated, while for small dephasing the dynamics can only be simulated in the presence of strong interactions (large $\abs{J_z}$).

\section{OE offset scaling} \label{sec:oeoffset}

Whether a state can be efficiently approximated by an MPO is determined by the OE of that state, which is at long times dominated by the logarithmic growth $S_\mathrm{OP} = \log_2(tJ)/4 + S_0$. Since logarithmic growth is very slow, the offset is also important. We find that the offset  $S_0$ decreases with increasing $|J_z|$ or $\gamma$  [see Fig.~\ref{fig:peakheights}(b)] when extrapolating to $t_0\to\infty$. The approximately equal spacing of lines in a regime of large $\gamma$ (when doubling $\gamma$) indicates a scaling of $S_0 \sim \log_2(1/\gamma)$ in this regime. This is confirmed by the perturbative argument given below, which leads to $S_0 = \log_2(J/2\sqrt{4\gamma^2+J_z^2})/4 + \mathcal O (1)$.

In the analytical model, the offsets are given by the time scale of the transfer, as the multiplicative transfer rate under the logarithm becomes an additive constant (see below). Cai and Barthel have computed the transfer rates perturbatively for small $J \ll \gamma$, as discussed in the following~\cite{cai2013algebraic}.

We start by writing the evolution operator as a sum $\mathcal L = \mathcal L_0 + \mathcal L_1$ with $\mathcal L_0 = -\mi [\hat H_0, \cdot ] + \sum_i \mathcal L^{[i]}$ and $\mathcal L_1 = - \mi [\hat H_1, \cdot]$. Here, the Hamiltonian is separated into interaction $\hat H_0 = J_z/4 \sum_i \hat \sigma_i^z \hat \sigma_{i+1}^z$ and $\hat H_1 = J/2 \sum_i \qty( \hat \sigma_i^+ \hat \sigma_{i+1}^- + \hat \sigma_i^- \hat \sigma_{i+1}^+)$. In the limit of strong dissipation, we can then consider perturbation theory for small $\mathcal L_1$.

The eigenstates of $\mathcal L_0$ are product states in the $z$ eigenbasis. Of these eigenstates, only those without off-diagonal correlations do not decay. Perturbative coupling between the states can be computed by [see also Eq.~\eqref{eq:strong_dephasing} in Sec.~\ref{sec:strongdephasing}]
\begin{align}
    \mathcal L_\mathrm{eff} = \mathcal L_0 - \mathcal P \mathcal L_1 \qty(\mathcal L_0)^{-1} \mathcal L_1 \mathcal P \, ,
\end{align}
where $\mathcal P$ projects into the $\mathcal L_0$ eigenstates with zero decay rate.

According to Ref.~\cite{cai2013algebraic}, the transfer rate between sites $i$ and $(i+1)$ depends on the state of the neighboring sites $(i-1)$ and $(i+2)$. Thus, focussing only on the two central spins, the transfer rates can be read from the prefactors of
\begin{align}
    \mathcal L_\mathrm{eff}^{i,(i+1)} \ket{\cdots \uparrow \uparrow \downarrow \uparrow\cdots}\bra{\cdots\uparrow \uparrow \downarrow \uparrow\cdots}
    &= \frac{J^2}{4\gamma} \big( \ket{\cdots\uparrow \downarrow \uparrow \uparrow\cdots}\bra{\cdots\uparrow \downarrow \uparrow \uparrow\cdots} - \ket{\cdots \uparrow \uparrow \downarrow \uparrow\cdots}\bra{\cdots\uparrow \uparrow \downarrow \uparrow\cdots} \big) \, , \\
    \mathcal L_\mathrm{eff}^{i,(i+1)} \ket{\cdots\uparrow \uparrow \downarrow \downarrow\cdots}\bra{\cdots\uparrow \uparrow \downarrow \downarrow\cdots}
    &= \frac{J^2\gamma}{4\gamma^2 + J_z^2} \big( \ket{\cdots\uparrow \downarrow\uparrow \downarrow\cdots}\bra{\cdots\uparrow \downarrow \uparrow \downarrow\cdots} - \ket{\cdots\uparrow \uparrow \downarrow \downarrow\cdots}\bra{\cdots\uparrow \uparrow \downarrow \downarrow\cdots} \big) \, .
\end{align}
All other possibilities can be obtained by permuting bras and kets, up and down, and reading states from right to left. Here, $\mathcal L_\mathrm{eff}^{i,(i+1)}$ indicates that only the effective Liouvillian acting on site $i$ and $i+1$ is given, i.e.~$\mathcal L_\mathrm{eff} = \sum_i \mathcal L_\mathrm{eff}^{i,(i+1)}$.

Since both scenarios are combinatorically equally likely for randomly arranged spins, we compute an effective transfer rate $r$ by taking the \emph{geometric} average, which yields
\begin{align}
    r = \frac{J^2}{2\sqrt{4\gamma^2+J_z^2}} \, .
\end{align}

From this, we can compute the parameter scaling of $S_0$ using $S(t) = \log_2(rt)/4 + \mathcal{O}(1)= \log_2(Jt)/4 + \log_2(r/J)/4  + \mathcal{O}(1)$ as
\begin{align}
    S_0 = \frac{1}{4}\log_2\qty[\frac{J}{2\sqrt{4\gamma^2+J_z^2}}]  + \mathcal{O}(1) \, .
\end{align}

\section{Strong dephasing limits of the Fermi-Hubbard model and of the Bose-Hubbard model}
\label{sec:strongdephasing}

\subsection{Strong dephasing limit of the Fermi-Hubbard model: two decoupled Symmetric Exclusion processes}

We define the Fermi-Hubbard Hamiltonian as
\begin{equation}
    \label{eq:FHham}
	\hat H =  -J \sum_{j=1}^N [ \hat c^\dagger_{\uparrow j+1} \hat c_{\uparrow j}  + \text{h.c.} + \hat c^\dagger_{\downarrow j+1} \hat c_{\downarrow j} + \text{h.c.} ] +  U\sum_{j=1}^N \hat c^\dagger_{\uparrow j} \hat c_{\uparrow j}   \hat c^\dagger_{\downarrow j} \hat c_{\downarrow j} ,
\end{equation}
where $\hat c^\dagger_{\uparrow j}$ (resp. $\hat c^\dagger_{\downarrow j}$) is the creation operator of a fermion with spin up (resp. down) on site $j$. We consider the strong dephasing limit of the Lindblad equation
\begin{equation}
	\label{eq:lindblad}
	\frac{d}{dt} \hat \rho  = - \mi [\hat H, \hat \rho] +  \gamma \sum_{j=1}^N \mathcal{D}[ \hat c^\dagger_{\uparrow j} \hat c_{\uparrow j} ] ( \hat \rho ) +  \gamma \sum_{j=1}^N \mathcal{D}[ \hat c^\dagger_{\downarrow j} \hat c_{\downarrow j} ] (\hat \rho ),
\end{equation} 
where $ \mathcal{D}[\hat A] (\rho) =  \hat A \hat \rho \hat A^\dagger  - \frac{1}{2} \{ \hat A^\dagger \hat A , \hat \rho \}  $. 
For our purposes, it is more convenient to write this equation in vectorized form: we vectorize the density matrix $\hat \rho \rightarrow \left| \rho \right>$ ---we write the $4^N \times 4^N$ matrix $\hat \rho$ as a $4^{2N}$-vector $\left| \rho \right>$---, and write its evolution as
\begin{equation}
	\frac{d}{dt} \left| \rho \right>  =   \gamma  \left(  \mathcal{L}_0 \left| \rho \right>   +  \frac{1}{\gamma} \mathcal{L}_1  \left| \rho \right> \right)
\end{equation}
where $\mathcal{L}_0$ contains the dissipative part of Eq.~(\ref{eq:lindblad}), and $\mathcal{L}_1$ contains the unitary part in~(\ref{eq:lindblad}). [Here, contrary to what is done in Ref.~\cite{cai2013algebraic} and in Sec.~\ref{sec:oeoffset}, we do not treat the diagonal and non-diagonal parts of the Hamiltonian differently. Here $\mathcal{L}_1$ corresponds to $- \mi [H,\cdot]$ where $H$ is the whole Hamiltonian, including its diagonal part.] Let $\tau = \gamma t$, then
\begin{equation}
	\frac{d}{d\tau} \left| \rho \right>  = \mathcal{L}_0 \left| \rho \right>   +  \frac{1}{\gamma} \mathcal{L}_1  \left| \rho \right> ,
\end{equation}
and the solution of that equation for some arbitrary initial condition $\left| \rho_0 \right>$ can be expanded to second order in $1/\gamma$,
\begin{eqnarray}
\nonumber	\left| \rho(\tau) \right>  &=&  \exp \left(  \tau \mathcal{L}_0  + \frac{\tau}{\gamma} \mathcal{L}_1  \right)  \left| \rho_0 \right> \\
\nonumber	& = & e^{ \tau \mathcal{L}_0 } \left| \rho_0 \right>  + \frac{1}{\gamma}  \int_0^\tau d\tau_1 e^{ (\tau - \tau_1) \mathcal{L}_0 }   \mathcal{L}_1 e^{ \tau_1 \mathcal{L}_0 } \left| \rho_0 \right> \\
	&&   +   \frac{1}{\gamma^2}  \int_0^\tau d\tau_1  \int_0^{\tau_1} d\tau_2 e^{  (\tau - \tau_1) \mathcal{L}_0 }   \mathcal{L}_1 e^{  (\tau_1-\tau_2) \mathcal{L}_0 }  \mathcal{L}_1 e^{ \tau_2 \mathcal{L}_0 }    \left| \rho_0 \right> + \dots
\end{eqnarray}
Importantly, the subspace of density matrices $\rho$ that satisfy $\mathcal{L}_0 \left| \rho \right> = 0$ is precisely the one of matrices that are diagonal in the computational basis. Let us call $\mathcal{P}$ the projector onto that subspace, and $\mathcal{P}^{\perp} = 1- \mathcal{P}$ its orthogonal projector. Then for $\tau \gg 1$ we can replace $e^{\tau \mathcal{L}_0}$ with $\mathcal{P}$. This leads to
\begin{eqnarray*}
	\left| \rho(\tau) \right>  &\simeq & \mathcal{P} \left| \rho_0 \right> + \frac{\tau}{\gamma}   \mathcal{P} \mathcal{L}_1 \mathcal{P} \left| \rho_0 \right> +   \frac{\tau^2}{2 \gamma^2}  \mathcal{P}   \mathcal{L}_1  \mathcal{P}   \mathcal{L}_1 \mathcal{P}    \left| \rho_0 \right> 
	 +   \frac{1}{\gamma^2} \int_0^{\tau} d\tau_2  (\tau-\tau_2)   \mathcal{P}   \mathcal{L}_1 \mathcal{P}^{\perp}  e^{ \tau_2 \mathcal{L}_0 }  \mathcal{P}^{\perp}  \mathcal{L}_1 \mathcal{P}    \left| \rho_0 \right>  + \dots \\
	 &\simeq & \exp \left(  \frac{\tau}{\gamma}  \mathcal{P}  \mathcal{L}_1  \mathcal{P}  \right)  \mathcal{P} \left| \rho_0 \right>  + \mathcal{P}   \mathcal{L}_1 \mathcal{P}^{\perp}  \left(  \frac{\tau}{\gamma^2} \int_0^{\infty} d\tau_2  e^{ \tau_2 \mathcal{L}_0 }  \right) \mathcal{P}^{\perp}  \mathcal{L}_1 \mathcal{P}    \left| \rho_0 \right> +\dots  \\
	&\simeq & \exp \left( \frac{\tau}{\gamma}  \mathcal{P}  \mathcal{L}_1  \mathcal{P}  \right)  \mathcal{P} \left| \rho_0 \right> +  \frac{\tau}{\gamma^2}  \mathcal{P} \mathcal{L}_1 \mathcal{P}^{\perp}  \left( -\mathcal{L}_0^\perp \right)^{-1} \mathcal{P}^{\perp}  \mathcal{L}_1 \mathcal{P}    \left| \rho_0 \right> +\dots ,
\end{eqnarray*}
where $\mathcal{L}_0^\perp$ is the restriction of $\mathcal{L}_0$ to the subspace orthogonal to its kernel, and $\left( \mathcal{L}_0^\perp \right)^{-1}$ is its inverse on that subspace. Finally, one notes that $\mathcal{P} \mathcal{L}_1 \mathcal{P} = 0$, since $\mathcal{L}_1$ acts on the density matrix $\rho$ as $-i [H,\rho]$. Thus, the result of second-order perturbation theory is that, in the strong dephasing limit, the density matrix  remains diagonal on long times scales, $\left| \rho(t) \right> \simeq \mathcal{P} \left| \rho(t) \right>$. The slow evolution of the diagonal part $ \left| \rho_{\rm diag} (t)\right> := \mathcal{P} \left| \rho(t) \right>$ is given by
\begin{equation}
	\label{eq:strong_dephasing}
	\frac{d}{dt} \left| \rho_{\rm diag}(t)\right>  \, = \, -\frac{1}{\gamma}  \mathcal{P}  \mathcal{L}_1 \mathcal{P}^{\perp} \left( \mathcal{L}_0^\perp \right)^{-1} \mathcal{P}^{\perp}  \mathcal{L}_1 \mathcal{P}     \left| \rho_{\rm diag}(t)\right> .
\end{equation}
This equation has been obtained for the XXZ chain by Cai and Barthel~\cite{cai2013algebraic} and also for other models, including models of bosons, by Bernard, Jin and Shpielberg~\cite{bernard2018transport}. Let us now apply it to the Fermi-Hubbard model (\ref{eq:FHham}). The key observation is that the interacting term $U \sum_{j=1}^N \hat c_{\uparrow j}^\dagger \hat c_{\uparrow j} \hat c_{\downarrow j}^\dagger \hat c_{\downarrow j}$ in the Hamiltonian (\ref{eq:FHham}) acts diagonally in the computational basis, so it does not contribute to the term $- \frac{1}{\gamma}  \mathcal{P}   \mathcal{L}_1 \mathcal{P}^{\perp}  \left( \mathcal{L}_0^\perp \right)^{-1} \mathcal{P}^{\perp}  \mathcal{L}_1 \mathcal{P} $ in Eq.~(\ref{eq:strong_dephasing}). Therefore, in the strong dephasing limit, the spin components $\uparrow$ and $\downarrow$ simply decouple, and each spin component follows its own strong-dephasing dynamics. One can then simply set $U=0$ in the Hamiltonian (\ref{eq:FHham}) which splits into two independent models of non-interacting fermions, one for each spin component. Consequently, the strong-dephasing limit of the Fermi-Hubbard model consists of two decoupled Symmetric Exclusion Processes, one for the spins $\uparrow$, the other for the spins $\downarrow$, as claimed in the main text.

\subsection{Strong dephasing limit of the Bose-Hubbard model: a Symmetric `Inclusion' Process}

We define the Bose-Hubbard Hamiltonian as
\begin{equation}
    \label{eq:BHham}
	\hat H = -J \sum_{j=1}^N [ \hat b^\dagger_{ j+1} \hat b_{j}  + \text{h.c.}  ] +  \frac{U}{2} \sum_{j=1}^N  \hat b^\dagger_{ j} \hat b_{j}   ( \hat b^\dagger_{ j} \hat b_{j} -1) 
\end{equation}
and the Lindlblad equation with dephasing takes the form (see e.g. Ref.~\cite{pichler2010nonequilibrium})
\begin{equation}
	\label{eq:lindblad_BH}
	\frac{d}{dt} \rho  = - \mi [\hat H, \hat \rho] +  \gamma \sum_{j=1}^N \mathcal{D}[ \hat b^\dagger_{j} \hat b_{j} ] ( \hat \rho ).
\end{equation}
Like for the XXZ chain and for the Fermi-Hubbard model, the strong dephasing limit can be analyzed in second-order perturbation theory~\cite{cai2013algebraic,bernard2018transport}. The general formula~(\ref{eq:strong_dephasing}), where $\gamma \mathcal{L}_0$ is the dissipative part of the Lindblad equation and $ \mathcal{L}_1$ is the unitary part, applies also to the Bose-Hubbard model. As in the Fermi-Hubbard case, the interaction term $\frac{U}{2} \sum_{j=1}^N  \hat b^\dagger_{ j} \hat b_{j}   ( \hat b^\dagger_{ j} \hat b_{j} -1) $ in the Hamiltonian (\ref{eq:BHham}) acts diagonally in the single-site Fock basis. Consequently, it does not contribute to the term $- \frac{1}{\gamma}  \mathcal{P}   \mathcal{L}_1 \mathcal{P}^{\perp}  \left( \mathcal{L}_0^\perp \right)^{-1} \mathcal{P}^{\perp}  \mathcal{L}_1 \mathcal{P} $. At strong dephasing the Hubbard interaction is irrelevant, and the model can be analyzed simply by setting $U=0$. The resulting model of non-interacting bosons with dephasing has been analyzed in Ref.~\cite{bernard2018transport}. For the convenience of the reader, here we briefly sketch a simple derivation of that strong-dephasing limit.

For notational simplicity, we focus on $N=2$ two sites only, and label the basis states as $\left| n_1 n_2 \right>$ where $n_1,n_2 \in \mathbb{N}$ are the number of bosons on site $1$ and $2$. Diagonal density matrices are of the form
\begin{equation}
 \rho_{\rm diag} = \sum_{n_1, n_2}  p(n_1,n_2)  \left| n_1 n_2 \right> \left< n_1 n_2 \right| .
\end{equation}
Under strong dephasing, the effective dynamics within the space of diagonal density matrices must be a classical stochastic process for the probability distribution $p(n_1,n_2)$. The goal is to determine the rates that define the classical master equation of that process.

We proceed step by step to compute the r.h.s of Eq.~(\ref{eq:strong_dephasing}). First, $ \mathcal{L}_1 \mathcal{P} \left| \rho_{\rm diag}\right> =  \mathcal{L}_1  \left| \rho_{\rm diag}\right>$ is the vectorized form of
\begin{eqnarray*}
  -\mi \left[ \hat H, \rho_{\rm diag} \right]  &=& -\mi \left[ H,  \sum_{n_1,n_2} p(n_1,n_2)  \left| n_1 n_2 \right> \left< n_1 n_2 \right| \right]  \\
 &=& \mi J \sum_{n_1, n_2}  p(n_1,n_2)  \left( \sqrt{(n_1+1) n_2}   \left| n_1+1, n_2-1 \right> \left< n_1 n_2 \right| + \sqrt{n_1 (n_2+1)}   \left| n_1-1, n_2+1 \right> \left< n_1 n_2 \right|  \right. \\
	&& - \left. \sqrt{(n_1+1) n_2}   \left| n_1, n_2 \right> \left< n_1+1 n_2-1 \right| + \sqrt{n_1 (n_2+1)}   \left| n_1, n_2 \right> \left< n_1-1 ,n_2 +1\right|  \right) ,
\end{eqnarray*}
which we write as
\begin{eqnarray*}
  \mathcal{L}_1 \left| \rho_{\rm diag} \right> &=& \mi J \sum_{n_1, n_2}  p(n_1,n_2)  \left( \sqrt{(n_1+1) n_2}   \left| n_1+1, n_2-1 \right> \left< n_1 n_2 \right| + \sqrt{n_1 (n_2+1)}   \left| n_1-1, n_2+1 \right> \left| n_1 n_2 \right>  \right. \\
	&& - \left. \sqrt{(n_1+1) n_2}   \left| n_1, n_2 \right> \left< n_1+1 n_2-1 \right| + \sqrt{n_1 (n_2+1)}   \left| n_1, n_2 \right> \left| n_1-1 ,n_2 +1\right>  \right) .
\end{eqnarray*}
Second, we observe that $\mathcal{L}[\hat b_1^\dagger \hat b_1] \left| n_1 \pm 1,  n_2' \right> \left| n_1, n_2 \right> = (n_1\pm 1) n_1 - \frac{(n_1\pm 1)^2 + n_1^2}{2} = - \frac{1}{2}$. Then
\begin{eqnarray*}
    (\mathcal{L}_0^\perp)^{-1} \mathcal{P}^\perp \mathcal{L}_1  \mathcal{P} \left| \rho_{\rm diag}\right>
	&=& \frac{-i J}{2 \gamma} \sum_{n_1, n_2}  p(n_1,n_2) 
	 \left( \sqrt{(n_1+1) n_2}   \left| n_1+1, n_2-1 \right> \left| n_1 n_2 \right> + \sqrt{n_1 (n_2+1)}   \left| n_1-1, n_2+1 \right> \left| n_1 n_2 \right>  \right. \\
	&& - \left. \sqrt{(n_1+1) n_2}   \left| n_1, n_2 \right> \left| n_1+1 n_2-1 \right> + \sqrt{n_1 (n_2+1)}   \left| n_1, n_2 \right> \left| n_1-1 ,n_2 +1\right>  \right) .
\end{eqnarray*}
Applying again the vectorized form of $- \mi [\hat H,.]$, and projecting onto diagonal configurations, one arrives at
\begin{eqnarray*}
	- \mathcal{P}\mathcal{L}_1 \mathcal{P}^\perp  (\mathcal{L}_0^\perp)^{-1} \mathcal{P}^\perp \mathcal{L}_1  \mathcal{P} \left| \rho_{\rm diag}\right> &=& \frac{J^2}{2\gamma} \sum_{n_1, n_2}  p(n_1,n_2)   \\
	&&  \left( \sqrt{(n_1+1) n_2}^2  (  \left| n_1+1, n_2-1 \right> \left| n_1+1 n_2-1 \right> - \left| n_1, n_2 \right> \left| n_1 n_2 \right> ) \right.   \\
	&&     + \sqrt{n_1 (n_2+1)}^2 (   \left| n_1-1, n_2+1 \right> \left| n_1-1 n_2+1 \right>  - \left| n_1, n_2 \right> \left| n_1 n_2 \right>   ) \\
	&&  +  \sqrt{(n_1+1) n_2}^2  (  \left| n_1+1, n_2-1 \right> \left| n_1+1 n_2-1 \right> - \left| n_1, n_2 \right> \left| n_1 n_2 \right> ) \\
	&& \left. + \sqrt{n_1 (n_2+1)}^2 (   \left| n_1-1, n_2+1 \right> \left| n_1-1 ,n_2 +1\right> -\left| n_1, n_2 \right> \left| n_1 ,n_2\right>  ) \right) .
\end{eqnarray*}
In conclusion, the classical master equation obeyed by the probability distribution $p(n_1, n_2)$ is
\begin{eqnarray}
\nonumber	\frac{d}{dt}  p(n_1,n_2) &=& \frac{J^2}{\gamma}  n_1 (n_2+1)  [ p(n_1-1,n_2+1)   -p(n_1,n_2)   ] \\
	 && +  \frac{J^2}{\gamma}  (n_1+1) n_2  [ p(n_1+1,n_2-1)   -p(n_1,n_2)   ] .
\end{eqnarray}
This result generalizes straightforwardly to the case with more sites, $N>2$, i.e.
\begin{align}
    \label{eq:masterbosons}
	\frac{d}{dt}  p(n_1,n_2, \dots, n_N) = \frac{J^2}{\gamma}   \sum_{i}  n_i  [ &(n_{i-1}+1)p( \dots, n_{i-1}+1, n_i-1, \dots) + (n_{i+1}+1)p( \dots,  n_i-1,n_{i+1}+1, \dots) \nonumber\\
	& -  (n_{i-1} + n_{i+1} + 2)p (  \dots,  n_i, n_{i+1}, \dots)  ] .
\end{align}

What is interesting about the master equation (\ref{eq:masterbosons}) is that it is very different from the one of non-interacting particles undergoing a random walk, which would rather be of the form
\begin{eqnarray*}
\nonumber	\frac{d}{dt}  p(n_1,n_2, \dots, n_N) & \propto &  \sum_{i}  n_i   [ p( \dots,  n_i-1, n_{i+1}+1, \dots) + p( \dots,  n_{i-1}+1, n_i-1, \dots) - 2p (  \dots, n_{i-1}, n_i, n_{i+1},\dots)  ] .
\end{eqnarray*}
Interestingly, the resulting classical dynamics exhibits a signature of bosonic bunching in the original quantum model.
The classical stochastic process defined by Eq.~(\ref{eq:masterbosons}) belongs to a class of models known as `Symmetric Inclusion Processes' in the statistical physics literature, see e.g. Refs.~\cite{giardina2010correlation,grosskinsky2013dynamics}.

\begin{figure}
    \centering
    \includegraphics[width=0.4\textwidth]{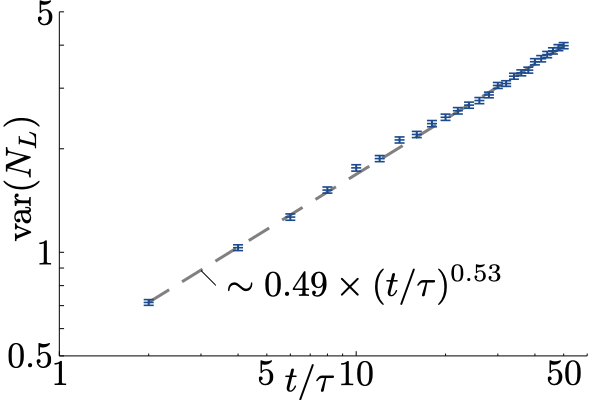}
    \caption{Fit for the variance growth for the Symmetric Inclusion Process. The blue crosses are determined by a Monte Carlo simulation of Eq.~\eqref{eq:masterbosons} with 5000 trajectories. The grey dashed line is a power law fit to the numerical data.}
    \label{fig:sip_fit}
\end{figure}

\bigskip

In order to determine the dynamics of number fluctuations in the model described by Eq.~\eqref{eq:masterbosons}, we numerically compute $n_\mathrm{samples}=5000$ different sample trajectories for a chain of length $L=100$. Fig.~\ref{fig:sip_fit} shows the time trace of the variance of the number of particles in the left half of the chain $\text{var}(N_L)$ [corrected by $n_\mathrm{samples}/(n_\mathrm{samples}-1)$]. We compute the error of the variance using the jackknife method~\cite{miller1974jackknife} by separating the data into $M=500$ bins with 10 trajectories each. Then, we compute the variances $v_m$ for all possible subsets of $M-1$ different bins (i.e.~deleting 1 bin for each subset). Finally, we compute the error from these variances as
\begin{align}
    \text{error}[\text{var}(N_L)] = \sqrt{\frac{M-1}{M} \sum_m \qty[v_m - \text{var}(N_L)]^2} \, .
\end{align}

We use \texttt{LsqFit.jl} to fit the numerical data with a power law of the form $\text{var}(N_L) = a \times (t/\tau)^b$, and find $a \approx 0.49$ and $b = 0.53$, close to $b=1/2$. We find that this fit describes the data well, as shown in Fig.~\ref{fig:sip_fit}. Here, $\tau=\gamma/J^2$ is the characteristic time scale of the dynamics.

If we thus assume that $N_L$ is Gaussian distributed with standard deviation $\delta \sim (t/\tau)^{b/2}$, we can use Eq.~(7) of the main text to compute $S_\mathrm{num}(t) = \log_2[(t/\tau)^{b/2}] + \mathcal O(1)$, close to $\log_2(t/\tau)/4 + \text{const.}$ used in the main text.

\end{widetext}

\bibliography{main}

\begin{thebibliography}{90}%
\makeatletter
\providecommand \@ifxundefined [1]{%
 \@ifx{#1\undefined}
}%
\providecommand \@ifnum [1]{%
 \ifnum #1\expandafter \@firstoftwo
 \else \expandafter \@secondoftwo
 \fi
}%
\providecommand \@ifx [1]{%
 \ifx #1\expandafter \@firstoftwo
 \else \expandafter \@secondoftwo
 \fi
}%
\providecommand \natexlab [1]{#1}%
\providecommand \enquote  [1]{``#1''}%
\providecommand \bibnamefont  [1]{#1}%
\providecommand \bibfnamefont [1]{#1}%
\providecommand \citenamefont [1]{#1}%
\providecommand \href@noop [0]{\@secondoftwo}%
\providecommand \href [0]{\begingroup \@sanitize@url \@href}%
\providecommand \@href[1]{\@@startlink{#1}\@@href}%
\providecommand \@@href[1]{\endgroup#1\@@endlink}%
\providecommand \@sanitize@url [0]{\catcode `\\12\catcode `\$12\catcode
  `\&12\catcode `\#12\catcode `\^12\catcode `\_12\catcode `\%12\relax}%
\providecommand \@@startlink[1]{}%
\providecommand \@@endlink[0]{}%
\providecommand \url  [0]{\begingroup\@sanitize@url \@url }%
\providecommand \@url [1]{\endgroup\@href {#1}{\urlprefix }}%
\providecommand \urlprefix  [0]{URL }%
\providecommand \Eprint [0]{\href }%
\providecommand \doibase [0]{https://doi.org/}%
\providecommand \selectlanguage [0]{\@gobble}%
\providecommand \bibinfo  [0]{\@secondoftwo}%
\providecommand \bibfield  [0]{\@secondoftwo}%
\providecommand \translation [1]{[#1]}%
\providecommand \BibitemOpen [0]{}%
\providecommand \bibitemStop [0]{}%
\providecommand \bibitemNoStop [0]{.\EOS\space}%
\providecommand \EOS [0]{\spacefactor3000\relax}%
\providecommand \BibitemShut  [1]{\csname bibitem#1\endcsname}%
\let\auto@bib@innerbib\@empty
\bibitem [{\citenamefont {Amico}\ \emph {et~al.}(2008)\citenamefont {Amico},
  \citenamefont {Fazio}, \citenamefont {Osterloh},\ and\ \citenamefont
  {Vedral}}]{amico2008entanglement}%
  \BibitemOpen
  \bibfield  {author} {\bibinfo {author} {\bibfnamefont {L.}~\bibnamefont
  {Amico}}, \bibinfo {author} {\bibfnamefont {R.}~\bibnamefont {Fazio}},
  \bibinfo {author} {\bibfnamefont {A.}~\bibnamefont {Osterloh}},\ and\
  \bibinfo {author} {\bibfnamefont {V.}~\bibnamefont {Vedral}},\ }\bibfield
  {title} {\bibinfo {title} {{Entanglement in many-body systems}},\ }\href
  {https://doi.org/10.1103/RevModPhys.80.517} {\bibfield  {journal} {\bibinfo
  {journal} {Rev. Mod. Phys.}\ }\textbf {\bibinfo {volume} {80}},\ \bibinfo
  {pages} {517} (\bibinfo {year} {2008})}\BibitemShut {NoStop}%
\bibitem [{\citenamefont {Eisert}\ \emph {et~al.}(2010)\citenamefont {Eisert},
  \citenamefont {Cramer},\ and\ \citenamefont {Plenio}}]{eisert2010colloquium}%
  \BibitemOpen
  \bibfield  {author} {\bibinfo {author} {\bibfnamefont {J.}~\bibnamefont
  {Eisert}}, \bibinfo {author} {\bibfnamefont {M.}~\bibnamefont {Cramer}},\
  and\ \bibinfo {author} {\bibfnamefont {M.~B.}\ \bibnamefont {Plenio}},\
  }\bibfield  {title} {\bibinfo {title} {{Colloquium: Area laws for the
  entanglement entropy}},\ }\href {https://doi.org/10.1103/RevModPhys.82.277}
  {\bibfield  {journal} {\bibinfo  {journal} {Rev. Mod. Phys.}\ }\textbf
  {\bibinfo {volume} {82}},\ \bibinfo {pages} {277} (\bibinfo {year}
  {2010})}\BibitemShut {NoStop}%
\bibitem [{\citenamefont {Calabrese}\ and\ \citenamefont
  {Cardy}(2005)}]{calabrese2005evolution}%
  \BibitemOpen
  \bibfield  {author} {\bibinfo {author} {\bibfnamefont {P.}~\bibnamefont
  {Calabrese}}\ and\ \bibinfo {author} {\bibfnamefont {J.}~\bibnamefont
  {Cardy}},\ }\bibfield  {title} {\bibinfo {title} {{Evolution of entanglement
  entropy in one-dimensional systems}},\ }\href
  {https://doi.org/10.1088/1742-5468/2005/04/p04010} {\bibfield  {journal}
  {\bibinfo  {journal} {J. Stat. Mech.: Theory Exp.}\ }\textbf {\bibinfo
  {volume} {2005}}\bibinfo  {number} { (04)},\ \bibinfo {pages}
  {P04010}}\BibitemShut {NoStop}%
\bibitem [{\citenamefont {Fagotti}\ and\ \citenamefont
  {Calabrese}(2008)}]{fagotti2008evolution}%
  \BibitemOpen
\bibfield  {number} {  }\bibfield  {author} {\bibinfo {author} {\bibfnamefont
  {M.}~\bibnamefont {Fagotti}}\ and\ \bibinfo {author} {\bibfnamefont
  {P.}~\bibnamefont {Calabrese}},\ }\bibfield  {title} {\bibinfo {title}
  {Evolution of entanglement entropy following a quantum quench: Analytic
  results for the $xy$ chain in a transverse magnetic field},\ }\href
  {https://doi.org/10.1103/PhysRevA.78.010306} {\bibfield  {journal} {\bibinfo
  {journal} {Phys. Rev. A}\ }\textbf {\bibinfo {volume} {78}},\ \bibinfo
  {pages} {010306(R)} (\bibinfo {year} {2008})}\BibitemShut {NoStop}%
\bibitem [{\citenamefont
  {{\ifmmode\check{Z}\else\v{Z}\fi}nidari{\ifmmode\check{c}\else\v{c}\fi}}\
  \emph {et~al.}(2008)\citenamefont
  {{\ifmmode\check{Z}\else\v{Z}\fi}nidari{\ifmmode\check{c}\else\v{c}\fi}},
  \citenamefont {Prosen},\ and\ \citenamefont
  {Prelov{\ifmmode\check{s}\else\v{s}\fi}ek}}]{znidaric2008many}%
  \BibitemOpen
  \bibfield  {author} {\bibinfo {author} {\bibfnamefont {M.}~\bibnamefont
  {{\ifmmode\check{Z}\else\v{Z}\fi}nidari{\ifmmode\check{c}\else\v{c}\fi}}},
  \bibinfo {author} {\bibfnamefont {T.}~\bibnamefont {Prosen}},\ and\ \bibinfo
  {author} {\bibfnamefont {P.}~\bibnamefont
  {Prelov{\ifmmode\check{s}\else\v{s}\fi}ek}},\ }\bibfield  {title} {\bibinfo
  {title} {{Many-body localization in the Heisenberg $XXZ$ magnet in a random
  field}},\ }\href {https://doi.org/10.1103/PhysRevB.77.064426} {\bibfield
  {journal} {\bibinfo  {journal} {Phys. Rev. B}\ }\textbf {\bibinfo {volume}
  {77}},\ \bibinfo {pages} {064426} (\bibinfo {year} {2008})}\BibitemShut
  {NoStop}%
\bibitem [{\citenamefont {Alba}\ and\ \citenamefont
  {Calabrese}(2017)}]{alba2017entanglement}%
  \BibitemOpen
  \bibfield  {author} {\bibinfo {author} {\bibfnamefont {V.}~\bibnamefont
  {Alba}}\ and\ \bibinfo {author} {\bibfnamefont {P.}~\bibnamefont
  {Calabrese}},\ }\bibfield  {title} {\bibinfo {title} {{Entanglement and
  thermodynamics after a quantum quench in integrable systems}},\ }\href
  {https://doi.org/10.1073/pnas.1703516114} {\bibfield  {journal} {\bibinfo
  {journal} {Proc. Natl. Acad. Sci. U. S. A.}\ }\textbf {\bibinfo {volume}
  {114}},\ \bibinfo {pages} {7947} (\bibinfo {year} {2017})}\BibitemShut
  {NoStop}%
\bibitem [{\citenamefont {Jonay}\ \emph {et~al.}(2018)\citenamefont {Jonay},
  \citenamefont {Huse},\ and\ \citenamefont {Nahum}}]{jonay2018coarse}%
  \BibitemOpen
  \bibfield  {author} {\bibinfo {author} {\bibfnamefont {C.}~\bibnamefont
  {Jonay}}, \bibinfo {author} {\bibfnamefont {D.~A.}\ \bibnamefont {Huse}},\
  and\ \bibinfo {author} {\bibfnamefont {A.}~\bibnamefont {Nahum}},\ }\bibfield
   {title} {\bibinfo {title} {Coarse-grained dynamics of operator and state
  entanglement},\ }\href {https://arxiv.org/abs/1803.00089} {\bibfield
  {journal} {\bibinfo  {journal} {arXiv:1803.00089}\ } (\bibinfo {year}
  {2018})}\BibitemShut {NoStop}%
\bibitem [{\citenamefont {Lukin}\ \emph {et~al.}(2019)\citenamefont {Lukin},
  \citenamefont {Rispoli}, \citenamefont {Schittko}, \citenamefont {Tai},
  \citenamefont {Kaufman}, \citenamefont {Choi}, \citenamefont {Khemani},
  \citenamefont {L{\ifmmode\acute{e}\else\'{e}\fi}onard},\ and\ \citenamefont
  {Greiner}}]{lukin2019probing}%
  \BibitemOpen
  \bibfield  {author} {\bibinfo {author} {\bibfnamefont {A.}~\bibnamefont
  {Lukin}}, \bibinfo {author} {\bibfnamefont {M.}~\bibnamefont {Rispoli}},
  \bibinfo {author} {\bibfnamefont {R.}~\bibnamefont {Schittko}}, \bibinfo
  {author} {\bibfnamefont {M.~E.}\ \bibnamefont {Tai}}, \bibinfo {author}
  {\bibfnamefont {A.~M.}\ \bibnamefont {Kaufman}}, \bibinfo {author}
  {\bibfnamefont {S.}~\bibnamefont {Choi}}, \bibinfo {author} {\bibfnamefont
  {V.}~\bibnamefont {Khemani}}, \bibinfo {author} {\bibfnamefont
  {J.}~\bibnamefont {L{\ifmmode\acute{e}\else\'{e}\fi}onard}},\ and\ \bibinfo
  {author} {\bibfnamefont {M.}~\bibnamefont {Greiner}},\ }\bibfield  {title}
  {\bibinfo {title} {{Probing entanglement in a many-body{\textendash}localized
  system}},\ }\href {https://www.science.org/doi/10.1126/science.aau0818}
  {\bibfield  {journal} {\bibinfo  {journal} {Science}\ }\textbf {\bibinfo
  {volume} {364}},\ \bibinfo {pages} {256} (\bibinfo {year}
  {2019})}\BibitemShut {NoStop}%
\bibitem [{\citenamefont {Vidal}(2004)}]{vidal2004efficient}%
  \BibitemOpen
  \bibfield  {author} {\bibinfo {author} {\bibfnamefont {G.}~\bibnamefont
  {Vidal}},\ }\bibfield  {title} {\bibinfo {title} {{Efficient Simulation of
  One-Dimensional Quantum Many-Body Systems}},\ }\href
  {https://doi.org/10.1103/PhysRevLett.93.040502} {\bibfield  {journal}
  {\bibinfo  {journal} {Phys. Rev. Lett.}\ }\textbf {\bibinfo {volume} {93}},\
  \bibinfo {pages} {040502} (\bibinfo {year} {2004})}\BibitemShut {NoStop}%
\bibitem [{\citenamefont {Verstraete}\ \emph {et~al.}(2008)\citenamefont
  {Verstraete}, \citenamefont {Murg},\ and\ \citenamefont
  {Cirac}}]{verstraete2008matrix}%
  \BibitemOpen
  \bibfield  {author} {\bibinfo {author} {\bibfnamefont {F.}~\bibnamefont
  {Verstraete}}, \bibinfo {author} {\bibfnamefont {V.}~\bibnamefont {Murg}},\
  and\ \bibinfo {author} {\bibfnamefont {J.~I.}\ \bibnamefont {Cirac}},\
  }\bibfield  {title} {\bibinfo {title} {{Matrix product states, projected
  entangled pair states, and variational renormalization group methods for
  quantum spin systems}},\ }\href {https://doi.org/10.1080/14789940801912366}
  {\bibfield  {journal} {\bibinfo  {journal} {Adv. Phys.}\ }\textbf {\bibinfo
  {volume} {57}},\ \bibinfo {pages} {143} (\bibinfo {year} {2008})}\BibitemShut
  {NoStop}%
\bibitem [{\citenamefont
  {Schollw{\ifmmode\ddot{o}\else\"{o}\fi}ck}(2011)}]{schollwock2011density-matrix}%
  \BibitemOpen
  \bibfield  {author} {\bibinfo {author} {\bibfnamefont {U.}~\bibnamefont
  {Schollw{\ifmmode\ddot{o}\else\"{o}\fi}ck}},\ }\bibfield  {title} {\bibinfo
  {title} {{The density-matrix renormalization group in the age of matrix
  product states}},\ }\href {https://doi.org/10.1016/j.aop.2010.09.012}
  {\bibfield  {journal} {\bibinfo  {journal} {Ann. Phys.}\ }\textbf {\bibinfo
  {volume} {326}},\ \bibinfo {pages} {96} (\bibinfo {year} {2011})}\BibitemShut
  {NoStop}%
\bibitem [{\citenamefont {Paeckel}\ \emph {et~al.}(2019)\citenamefont
  {Paeckel}, \citenamefont {K{\ifmmode\ddot{o}\else\"{o}\fi}hler},
  \citenamefont {Swoboda}, \citenamefont {Manmana}, \citenamefont
  {Schollw{\ifmmode\ddot{o}\else\"{o}\fi}ck},\ and\ \citenamefont
  {Hubig}}]{paeckel2019time-evolution}%
  \BibitemOpen
  \bibfield  {author} {\bibinfo {author} {\bibfnamefont {S.}~\bibnamefont
  {Paeckel}}, \bibinfo {author} {\bibfnamefont {T.}~\bibnamefont
  {K{\ifmmode\ddot{o}\else\"{o}\fi}hler}}, \bibinfo {author} {\bibfnamefont
  {A.}~\bibnamefont {Swoboda}}, \bibinfo {author} {\bibfnamefont {S.~R.}\
  \bibnamefont {Manmana}}, \bibinfo {author} {\bibfnamefont {U.}~\bibnamefont
  {Schollw{\ifmmode\ddot{o}\else\"{o}\fi}ck}},\ and\ \bibinfo {author}
  {\bibfnamefont {C.}~\bibnamefont {Hubig}},\ }\bibfield  {title} {\bibinfo
  {title} {{Time-evolution methods for matrix-product states}},\ }\href
  {https://doi.org/10.1016/j.aop.2019.167998} {\bibfield  {journal} {\bibinfo
  {journal} {Ann. Phys.}\ }\textbf {\bibinfo {volume} {411}},\ \bibinfo {pages}
  {167998} (\bibinfo {year} {2019})}\BibitemShut {NoStop}%
\bibitem [{\citenamefont {Schuch}\ \emph {et~al.}(2008)\citenamefont {Schuch},
  \citenamefont {Wolf}, \citenamefont {Verstraete},\ and\ \citenamefont
  {Cirac}}]{schuch2008entropy}%
  \BibitemOpen
  \bibfield  {author} {\bibinfo {author} {\bibfnamefont {N.}~\bibnamefont
  {Schuch}}, \bibinfo {author} {\bibfnamefont {M.~M.}\ \bibnamefont {Wolf}},
  \bibinfo {author} {\bibfnamefont {F.}~\bibnamefont {Verstraete}},\ and\
  \bibinfo {author} {\bibfnamefont {J.~I.}\ \bibnamefont {Cirac}},\ }\bibfield
  {title} {\bibinfo {title} {{Entropy Scaling and Simulability by Matrix
  Product States}},\ }\href {https://doi.org/10.1103/PhysRevLett.100.030504}
  {\bibfield  {journal} {\bibinfo  {journal} {Phys. Rev. Lett.}\ }\textbf
  {\bibinfo {volume} {100}},\ \bibinfo {pages} {030504} (\bibinfo {year}
  {2008})}\BibitemShut {NoStop}%
\bibitem [{\citenamefont {Bloch}\ \emph {et~al.}(2012)\citenamefont {Bloch},
  \citenamefont {Dalibard},\ and\ \citenamefont
  {Nascimb{\ifmmode\grave{e}\else\`{e}\fi}ne}}]{bloch2012quant}%
  \BibitemOpen
  \bibfield  {author} {\bibinfo {author} {\bibfnamefont {I.}~\bibnamefont
  {Bloch}}, \bibinfo {author} {\bibfnamefont {J.}~\bibnamefont {Dalibard}},\
  and\ \bibinfo {author} {\bibfnamefont {S.}~\bibnamefont
  {Nascimb{\ifmmode\grave{e}\else\`{e}\fi}ne}},\ }\bibfield  {title} {\bibinfo
  {title} {{Quantum simulations with ultracold quantum gases}},\ }\href
  {https://doi.org/10.1038/nphys2259} {\bibfield  {journal} {\bibinfo
  {journal} {Nat. Phys.}\ }\textbf {\bibinfo {volume} {8}},\ \bibinfo {pages}
  {267} (\bibinfo {year} {2012})}\BibitemShut {NoStop}%
\bibitem [{\citenamefont {Adams}\ \emph {et~al.}(2019)\citenamefont {Adams},
  \citenamefont {Pritchard},\ and\ \citenamefont {Shaffer}}]{adams2019rydberg}%
  \BibitemOpen
  \bibfield  {author} {\bibinfo {author} {\bibfnamefont {C.~S.}\ \bibnamefont
  {Adams}}, \bibinfo {author} {\bibfnamefont {J.~D.}\ \bibnamefont
  {Pritchard}},\ and\ \bibinfo {author} {\bibfnamefont {J.~P.}\ \bibnamefont
  {Shaffer}},\ }\bibfield  {title} {\bibinfo {title} {{Rydberg atom quantum
  technologies}},\ }\href {https://doi.org/10.1088/1361-6455/ab52ef} {\bibfield
   {journal} {\bibinfo  {journal} {J. Phys. B: At. Mol. Opt. Phys.}\ }\textbf
  {\bibinfo {volume} {53}},\ \bibinfo {pages} {012002} (\bibinfo {year}
  {2019})}\BibitemShut {NoStop}%
\bibitem [{\citenamefont {Browaeys}\ and\ \citenamefont
  {Lahaye}(2020)}]{browaeys2020many}%
  \BibitemOpen
  \bibfield  {author} {\bibinfo {author} {\bibfnamefont {A.}~\bibnamefont
  {Browaeys}}\ and\ \bibinfo {author} {\bibfnamefont {T.}~\bibnamefont
  {Lahaye}},\ }\bibfield  {title} {\bibinfo {title} {{Many-body physics with
  individually controlled Rydberg atoms}},\ }\href
  {https://doi.org/10.1038/s41567-019-0733-z} {\bibfield  {journal} {\bibinfo
  {journal} {Nat. Phys.}\ }\textbf {\bibinfo {volume} {16}},\ \bibinfo {pages}
  {132} (\bibinfo {year} {2020})}\BibitemShut {NoStop}%
\bibitem [{\citenamefont {Morgado}\ and\ \citenamefont
  {Whitlock}(2021)}]{morgado2021quantum}%
  \BibitemOpen
  \bibfield  {author} {\bibinfo {author} {\bibfnamefont {M.}~\bibnamefont
  {Morgado}}\ and\ \bibinfo {author} {\bibfnamefont {S.}~\bibnamefont
  {Whitlock}},\ }\bibfield  {title} {\bibinfo {title} {{Quantum simulation and
  computing with Rydberg-interacting qubits}},\ }\href
  {https://doi.org/10.1116/5.0036562} {\bibfield  {journal} {\bibinfo
  {journal} {AVS Quantum Sci.}\ }\textbf {\bibinfo {volume} {3}},\ \bibinfo
  {pages} {023501} (\bibinfo {year} {2021})}\BibitemShut {NoStop}%
\bibitem [{\citenamefont {Gadway}\ and\ \citenamefont
  {Yan}(2016)}]{gadway2016strong}%
  \BibitemOpen
  \bibfield  {author} {\bibinfo {author} {\bibfnamefont {B.}~\bibnamefont
  {Gadway}}\ and\ \bibinfo {author} {\bibfnamefont {B.}~\bibnamefont {Yan}},\
  }\bibfield  {title} {\bibinfo {title} {{Strongly interacting ultracold polar
  molecules}},\ }\href {https://doi.org/10.1088/0953-4075/49/15/152002}
  {\bibfield  {journal} {\bibinfo  {journal} {J. Phys. B: At. Mol. Opt. Phys.}\
  }\textbf {\bibinfo {volume} {49}},\ \bibinfo {pages} {152002} (\bibinfo
  {year} {2016})}\BibitemShut {NoStop}%
\bibitem [{\citenamefont {Blatt}\ and\ \citenamefont
  {Roos}(2012)}]{blatt2012quant}%
  \BibitemOpen
  \bibfield  {author} {\bibinfo {author} {\bibfnamefont {R.}~\bibnamefont
  {Blatt}}\ and\ \bibinfo {author} {\bibfnamefont {C.~F.}\ \bibnamefont
  {Roos}},\ }\bibfield  {title} {\bibinfo {title} {{Quantum simulations with
  trapped ions}},\ }\href {https://doi.org/10.1038/nphys2252} {\bibfield
  {journal} {\bibinfo  {journal} {Nat. Phys.}\ }\textbf {\bibinfo {volume}
  {8}},\ \bibinfo {pages} {277} (\bibinfo {year} {2012})}\BibitemShut {NoStop}%
\bibitem [{\citenamefont {Cirac}\ and\ \citenamefont
  {Zoller}(2012)}]{cirac2012goals}%
  \BibitemOpen
  \bibfield  {author} {\bibinfo {author} {\bibfnamefont {J.~I.}\ \bibnamefont
  {Cirac}}\ and\ \bibinfo {author} {\bibfnamefont {P.}~\bibnamefont {Zoller}},\
  }\bibfield  {title} {\bibinfo {title} {{Goals and opportunities in quantum
  simulation}},\ }\href {https://doi.org/10.1038/nphys2275} {\bibfield
  {journal} {\bibinfo  {journal} {Nat. Phys.}\ }\textbf {\bibinfo {volume}
  {8}},\ \bibinfo {pages} {264} (\bibinfo {year} {2012})}\BibitemShut {NoStop}%
\bibitem [{\citenamefont {Georgescu}\ \emph {et~al.}(2014)\citenamefont
  {Georgescu}, \citenamefont {Ashhab},\ and\ \citenamefont
  {Nori}}]{georgescu2014quantum}%
  \BibitemOpen
  \bibfield  {author} {\bibinfo {author} {\bibfnamefont {I.~M.}\ \bibnamefont
  {Georgescu}}, \bibinfo {author} {\bibfnamefont {S.}~\bibnamefont {Ashhab}},\
  and\ \bibinfo {author} {\bibfnamefont {F.}~\bibnamefont {Nori}},\ }\bibfield
  {title} {\bibinfo {title} {{Quantum simulation}},\ }\href
  {https://doi.org/10.1103/RevModPhys.86.153} {\bibfield  {journal} {\bibinfo
  {journal} {Rev. Mod. Phys.}\ }\textbf {\bibinfo {volume} {86}},\ \bibinfo
  {pages} {153} (\bibinfo {year} {2014})}\BibitemShut {NoStop}%
\bibitem [{\citenamefont {Zanardi}\ \emph {et~al.}(2000)\citenamefont
  {Zanardi}, \citenamefont {Zalka},\ and\ \citenamefont
  {Faoro}}]{zanardi2000entangling}%
  \BibitemOpen
  \bibfield  {author} {\bibinfo {author} {\bibfnamefont {P.}~\bibnamefont
  {Zanardi}}, \bibinfo {author} {\bibfnamefont {C.}~\bibnamefont {Zalka}},\
  and\ \bibinfo {author} {\bibfnamefont {L.}~\bibnamefont {Faoro}},\ }\bibfield
   {title} {\bibinfo {title} {{Entangling power of quantum evolutions}},\
  }\href {https://doi.org/10.1103/PhysRevA.62.030301} {\bibfield  {journal}
  {\bibinfo  {journal} {Phys. Rev. A}\ }\textbf {\bibinfo {volume} {62}},\
  \bibinfo {pages} {030301(R)} (\bibinfo {year} {2000})}\BibitemShut {NoStop}%
\bibitem [{\citenamefont {Zanardi}(2001)}]{zanardi2001entanglement}%
  \BibitemOpen
  \bibfield  {author} {\bibinfo {author} {\bibfnamefont {P.}~\bibnamefont
  {Zanardi}},\ }\bibfield  {title} {\bibinfo {title} {{Entanglement of quantum
  evolutions}},\ }\href {https://doi.org/10.1103/PhysRevA.63.040304} {\bibfield
   {journal} {\bibinfo  {journal} {Phys. Rev. A}\ }\textbf {\bibinfo {volume}
  {63}},\ \bibinfo {pages} {040304(R)} (\bibinfo {year} {2001})}\BibitemShut
  {NoStop}%
\bibitem [{\citenamefont {Wang}\ and\ \citenamefont
  {Zanardi}(2002)}]{wang2002quantum}%
  \BibitemOpen
  \bibfield  {author} {\bibinfo {author} {\bibfnamefont {X.}~\bibnamefont
  {Wang}}\ and\ \bibinfo {author} {\bibfnamefont {P.}~\bibnamefont {Zanardi}},\
  }\bibfield  {title} {\bibinfo {title} {{Quantum entanglement of unitary
  operators on bipartite systems}},\ }\href
  {https://doi.org/10.1103/PhysRevA.66.044303} {\bibfield  {journal} {\bibinfo
  {journal} {Phys. Rev. A}\ }\textbf {\bibinfo {volume} {66}},\ \bibinfo
  {pages} {044303} (\bibinfo {year} {2002})}\BibitemShut {NoStop}%
\bibitem [{\citenamefont {Prosen}\ and\ \citenamefont
  {Pi{\ifmmode\check{z}\else\v{z}\fi}orn}(2007)}]{prosen2007operator}%
  \BibitemOpen
  \bibfield  {author} {\bibinfo {author} {\bibfnamefont {T.}~\bibnamefont
  {Prosen}}\ and\ \bibinfo {author} {\bibfnamefont {I.}~\bibnamefont
  {Pi{\ifmmode\check{z}\else\v{z}\fi}orn}},\ }\bibfield  {title} {\bibinfo
  {title} {{Operator space entanglement entropy in a transverse Ising chain}},\
  }\href {https://doi.org/10.1103/PhysRevA.76.032316} {\bibfield  {journal}
  {\bibinfo  {journal} {Phys. Rev. A}\ }\textbf {\bibinfo {volume} {76}},\
  \bibinfo {pages} {032316} (\bibinfo {year} {2007})}\BibitemShut {NoStop}%
\bibitem [{\citenamefont {Dubail}(2017)}]{dubail2017entanglement}%
  \BibitemOpen
  \bibfield  {author} {\bibinfo {author} {\bibfnamefont {J.}~\bibnamefont
  {Dubail}},\ }\bibfield  {title} {\bibinfo {title} {Entanglement scaling of
  operators: a conformal field theory approach, with a glimpse of simulability
  of long-time dynamics in 1+1d},\ }\href
  {https://doi.org/10.1088/1751-8121/aa6f38} {\bibfield  {journal} {\bibinfo
  {journal} {Journal of Physics A: Mathematical and Theoretical}\ }\textbf
  {\bibinfo {volume} {50}},\ \bibinfo {pages} {234001} (\bibinfo {year}
  {2017})}\BibitemShut {NoStop}%
\bibitem [{\citenamefont {Zhou}\ and\ \citenamefont
  {Luitz}(2017)}]{zhou2017operator}%
  \BibitemOpen
  \bibfield  {author} {\bibinfo {author} {\bibfnamefont {T.}~\bibnamefont
  {Zhou}}\ and\ \bibinfo {author} {\bibfnamefont {D.~J.}\ \bibnamefont
  {Luitz}},\ }\bibfield  {title} {\bibinfo {title} {{Operator entanglement
  entropy of the time evolution operator in chaotic systems}},\ }\href
  {https://doi.org/10.1103/PhysRevB.95.094206} {\bibfield  {journal} {\bibinfo
  {journal} {Phys. Rev. B}\ }\textbf {\bibinfo {volume} {95}},\ \bibinfo
  {pages} {094206} (\bibinfo {year} {2017})}\BibitemShut {NoStop}%
\bibitem [{\citenamefont {Alba}\ \emph {et~al.}(2019)\citenamefont {Alba},
  \citenamefont {Dubail},\ and\ \citenamefont {Medenjak}}]{alba2019operator}%
  \BibitemOpen
  \bibfield  {author} {\bibinfo {author} {\bibfnamefont {V.}~\bibnamefont
  {Alba}}, \bibinfo {author} {\bibfnamefont {J.}~\bibnamefont {Dubail}},\ and\
  \bibinfo {author} {\bibfnamefont {M.}~\bibnamefont {Medenjak}},\ }\bibfield
  {title} {\bibinfo {title} {Operator entanglement in interacting integrable
  quantum systems: The case of the rule 54 chain},\ }\href
  {https://doi.org/10.1103/PhysRevLett.122.250603} {\bibfield  {journal}
  {\bibinfo  {journal} {Phys. Rev. Lett.}\ }\textbf {\bibinfo {volume} {122}},\
  \bibinfo {pages} {250603} (\bibinfo {year} {2019})}\BibitemShut {NoStop}%
\bibitem [{\citenamefont {Wang}\ and\ \citenamefont
  {Zhou}(2019)}]{wang2019barrier}%
  \BibitemOpen
  \bibfield  {author} {\bibinfo {author} {\bibfnamefont {H.}~\bibnamefont
  {Wang}}\ and\ \bibinfo {author} {\bibfnamefont {T.}~\bibnamefont {Zhou}},\
  }\bibfield  {title} {\bibinfo {title} {{Barrier from chaos: operator
  entanglement dynamics of the reduced density matrix}},\ }\href
  {https://doi.org/10.1007/JHEP12(2019)020} {\bibfield  {journal} {\bibinfo
  {journal} {J. High Energy Phys.}\ }\textbf {\bibinfo {volume} {2019}}\bibinfo
   {number} { (12)},\ \bibinfo {pages} {1}}\BibitemShut {NoStop}%
\bibitem [{\citenamefont {Styliaris}\ \emph {et~al.}(2021)\citenamefont
  {Styliaris}, \citenamefont {Anand},\ and\ \citenamefont
  {Zanardi}}]{styliaris2021information}%
  \BibitemOpen
\bibfield  {number} {  }\bibfield  {author} {\bibinfo {author} {\bibfnamefont
  {G.}~\bibnamefont {Styliaris}}, \bibinfo {author} {\bibfnamefont
  {N.}~\bibnamefont {Anand}},\ and\ \bibinfo {author} {\bibfnamefont
  {P.}~\bibnamefont {Zanardi}},\ }\bibfield  {title} {\bibinfo {title}
  {{Information Scrambling over Bipartitions: Equilibration, Entropy
  Production, and Typicality}},\ }\href
  {https://doi.org/10.1103/PhysRevLett.126.030601} {\bibfield  {journal}
  {\bibinfo  {journal} {Phys. Rev. Lett.}\ }\textbf {\bibinfo {volume} {126}},\
  \bibinfo {pages} {030601} (\bibinfo {year} {2021})}\BibitemShut {NoStop}%
\bibitem [{Note1()}]{Note1}%
  \BibitemOpen
  \bibinfo {note} {Note that here this indicates approximability w.r.t.~the
  two-norm of the vectorized density matrix.}\BibitemShut {Stop}%
\bibitem [{\citenamefont {Noh}\ \emph {et~al.}(2020)\citenamefont {Noh},
  \citenamefont {Jiang},\ and\ \citenamefont {Fefferman}}]{noh2020efficient}%
  \BibitemOpen
  \bibfield  {author} {\bibinfo {author} {\bibfnamefont {K.}~\bibnamefont
  {Noh}}, \bibinfo {author} {\bibfnamefont {L.}~\bibnamefont {Jiang}},\ and\
  \bibinfo {author} {\bibfnamefont {B.}~\bibnamefont {Fefferman}},\ }\bibfield
  {title} {\bibinfo {title} {{Efficient classical simulation of noisy random
  quantum circuits in one dimension}},\ }\href
  {https://doi.org/10.22331/q-2020-09-11-318} {\bibfield  {journal} {\bibinfo
  {journal} {Quantum}\ }\textbf {\bibinfo {volume} {4}},\ \bibinfo {pages}
  {318} (\bibinfo {year} {2020})},\ \Eprint
  {https://arxiv.org/abs/2003.13163v3} {2003.13163v3} \BibitemShut {NoStop}%
\bibitem [{\citenamefont {Rakovszky}\ \emph {et~al.}(2020)\citenamefont
  {Rakovszky}, \citenamefont {von Keyserlingk},\ and\ \citenamefont
  {Pollmann}}]{rakovszky2020dissipation}%
  \BibitemOpen
  \bibfield  {author} {\bibinfo {author} {\bibfnamefont {T.}~\bibnamefont
  {Rakovszky}}, \bibinfo {author} {\bibfnamefont {C.~W.}\ \bibnamefont {von
  Keyserlingk}},\ and\ \bibinfo {author} {\bibfnamefont {F.}~\bibnamefont
  {Pollmann}},\ }\bibfield  {title} {\bibinfo {title} {{Dissipation-assisted
  operator evolution method for capturing hydrodynamic transport}},\ }\href
  {https://arxiv.org/abs/2004.05177v1} {\bibfield  {journal} {\bibinfo
  {journal} {arXiv:2004.05177}\ } (\bibinfo {year} {2020})}\BibitemShut
  {NoStop}%
\bibitem [{\citenamefont {Bertini}\ \emph
  {et~al.}(2020{\natexlab{a}})\citenamefont {Bertini}, \citenamefont {Kos},\
  and\ \citenamefont {Prosen}}]{bertini2020operatorI}%
  \BibitemOpen
  \bibfield  {author} {\bibinfo {author} {\bibfnamefont {B.}~\bibnamefont
  {Bertini}}, \bibinfo {author} {\bibfnamefont {P.}~\bibnamefont {Kos}},\ and\
  \bibinfo {author} {\bibfnamefont {T.}~\bibnamefont {Prosen}},\ }\bibfield
  {title} {\bibinfo {title} {{Operator Entanglement in Local Quantum Circuits
  I: Chaotic Dual-Unitary Circuits}},\ }\href
  {https://doi.org/10.21468/SciPostPhys.8.4.067} {\bibfield  {journal}
  {\bibinfo  {journal} {SciPost Phys.}\ }\textbf {\bibinfo {volume} {8}},\
  \bibinfo {pages} {067} (\bibinfo {year} {2020}{\natexlab{a}})}\BibitemShut
  {NoStop}%
\bibitem [{\citenamefont {Bertini}\ \emph
  {et~al.}(2020{\natexlab{b}})\citenamefont {Bertini}, \citenamefont {Kos},\
  and\ \citenamefont {Prosen}}]{bertini2020operatorII}%
  \BibitemOpen
  \bibfield  {author} {\bibinfo {author} {\bibfnamefont {B.}~\bibnamefont
  {Bertini}}, \bibinfo {author} {\bibfnamefont {P.}~\bibnamefont {Kos}},\ and\
  \bibinfo {author} {\bibfnamefont {T.}~\bibnamefont {Prosen}},\ }\bibfield
  {title} {\bibinfo {title} {{Operator Entanglement in Local Quantum Circuits
  II: Solitons in Chains of Qubits}},\ }\href
  {https://doi.org/10.21468/SciPostPhys.8.4.068} {\bibfield  {journal}
  {\bibinfo  {journal} {SciPost Phys.}\ }\textbf {\bibinfo {volume} {8}},\
  \bibinfo {pages} {068} (\bibinfo {year} {2020}{\natexlab{b}})}\BibitemShut
  {NoStop}%
\bibitem [{\citenamefont {Rossini}\ and\ \citenamefont
  {Vicari}(2021)}]{rossini2021coherent}%
  \BibitemOpen
  \bibfield  {author} {\bibinfo {author} {\bibfnamefont {D.}~\bibnamefont
  {Rossini}}\ and\ \bibinfo {author} {\bibfnamefont {E.}~\bibnamefont
  {Vicari}},\ }\bibfield  {title} {\bibinfo {title} {{Coherent and dissipative
  dynamics at quantum phase transitions}},\ }\href
  {https://doi.org/https://doi.org/10.1016/j.physrep.2021.08.003} {\bibfield
  {journal} {\bibinfo  {journal} {Physics Reports}\ }\textbf {\bibinfo {volume}
  {936}},\ \bibinfo {pages} {1} (\bibinfo {year} {2021})}\BibitemShut {NoStop}%
\bibitem [{\citenamefont {Cai}\ and\ \citenamefont
  {Barthel}(2013)}]{cai2013algebraic}%
  \BibitemOpen
  \bibfield  {author} {\bibinfo {author} {\bibfnamefont {Z.}~\bibnamefont
  {Cai}}\ and\ \bibinfo {author} {\bibfnamefont {T.}~\bibnamefont {Barthel}},\
  }\bibfield  {title} {\bibinfo {title} {{Algebraic versus Exponential
  Decoherence in Dissipative Many-Particle Systems}},\ }\href
  {https://doi.org/10.1103/PhysRevLett.111.150403} {\bibfield  {journal}
  {\bibinfo  {journal} {Phys. Rev. Lett.}\ }\textbf {\bibinfo {volume} {111}},\
  \bibinfo {pages} {150403} (\bibinfo {year} {2013})}\BibitemShut {NoStop}%
\bibitem [{\citenamefont {Medvedyeva}\ \emph
  {et~al.}(2016{\natexlab{a}})\citenamefont {Medvedyeva}, \citenamefont
  {Essler},\ and\ \citenamefont {Prosen}}]{medvedyeva2016exact}%
  \BibitemOpen
  \bibfield  {author} {\bibinfo {author} {\bibfnamefont {M.~V.}\ \bibnamefont
  {Medvedyeva}}, \bibinfo {author} {\bibfnamefont {F.~H.~L.}\ \bibnamefont
  {Essler}},\ and\ \bibinfo {author} {\bibfnamefont {T.}~\bibnamefont
  {Prosen}},\ }\bibfield  {title} {\bibinfo {title} {{Exact Bethe Ansatz
  Spectrum of a Tight-Binding Chain with Dephasing Noise}},\ }\href
  {https://doi.org/10.1103/PhysRevLett.117.137202} {\bibfield  {journal}
  {\bibinfo  {journal} {Phys. Rev. Lett.}\ }\textbf {\bibinfo {volume} {117}},\
  \bibinfo {pages} {137202} (\bibinfo {year} {2016}{\natexlab{a}})}\BibitemShut
  {NoStop}%
\bibitem [{\citenamefont {Foss-Feig}\ \emph {et~al.}(2017)\citenamefont
  {Foss-Feig}, \citenamefont {Young}, \citenamefont {Albert}, \citenamefont
  {Gorshkov},\ and\ \citenamefont {Maghrebi}}]{foss2017solvable}%
  \BibitemOpen
  \bibfield  {author} {\bibinfo {author} {\bibfnamefont {M.}~\bibnamefont
  {Foss-Feig}}, \bibinfo {author} {\bibfnamefont {J.~T.}\ \bibnamefont
  {Young}}, \bibinfo {author} {\bibfnamefont {V.~V.}\ \bibnamefont {Albert}},
  \bibinfo {author} {\bibfnamefont {A.~V.}\ \bibnamefont {Gorshkov}},\ and\
  \bibinfo {author} {\bibfnamefont {M.~F.}\ \bibnamefont {Maghrebi}},\
  }\bibfield  {title} {\bibinfo {title} {{Solvable Family of Driven-Dissipative
  Many-Body Systems}},\ }\href {https://doi.org/10.1103/PhysRevLett.119.190402}
  {\bibfield  {journal} {\bibinfo  {journal} {Phys. Rev. Lett.}\ }\textbf
  {\bibinfo {volume} {119}},\ \bibinfo {pages} {190402} (\bibinfo {year}
  {2017})}\BibitemShut {NoStop}%
\bibitem [{\citenamefont
  {{\ifmmode\check{Z}\else\v{Z}\fi}nidari{\ifmmode\check{c}\else\v{c}\fi}}(2015)}]{znidaric2015relaxation}%
  \BibitemOpen
  \bibfield  {author} {\bibinfo {author} {\bibfnamefont {M.}~\bibnamefont
  {{\ifmmode\check{Z}\else\v{Z}\fi}nidari{\ifmmode\check{c}\else\v{c}\fi}}},\
  }\bibfield  {title} {\bibinfo {title} {{Relaxation times of dissipative
  many-body quantum systems}},\ }\href
  {https://doi.org/10.1103/PhysRevE.92.042143} {\bibfield  {journal} {\bibinfo
  {journal} {Phys. Rev. E}\ }\textbf {\bibinfo {volume} {92}},\ \bibinfo
  {pages} {042143} (\bibinfo {year} {2015})}\BibitemShut {NoStop}%
\bibitem [{\citenamefont {Poletti}\ \emph {et~al.}(2012)\citenamefont
  {Poletti}, \citenamefont {Bernier}, \citenamefont {Georges},\ and\
  \citenamefont {Kollath}}]{poletti2012interaction}%
  \BibitemOpen
  \bibfield  {author} {\bibinfo {author} {\bibfnamefont {D.}~\bibnamefont
  {Poletti}}, \bibinfo {author} {\bibfnamefont {J.-S.}\ \bibnamefont
  {Bernier}}, \bibinfo {author} {\bibfnamefont {A.}~\bibnamefont {Georges}},\
  and\ \bibinfo {author} {\bibfnamefont {C.}~\bibnamefont {Kollath}},\
  }\bibfield  {title} {\bibinfo {title} {{Interaction-Induced Impeding of
  Decoherence and Anomalous Diffusion}},\ }\href
  {https://doi.org/10.1103/PhysRevLett.109.045302} {\bibfield  {journal}
  {\bibinfo  {journal} {Phys. Rev. Lett.}\ }\textbf {\bibinfo {volume} {109}},\
  \bibinfo {pages} {045302} (\bibinfo {year} {2012})}\BibitemShut {NoStop}%
\bibitem [{\citenamefont {Poletti}\ \emph {et~al.}(2013)\citenamefont
  {Poletti}, \citenamefont {Barmettler}, \citenamefont {Georges},\ and\
  \citenamefont {Kollath}}]{poletti2013emergence}%
  \BibitemOpen
  \bibfield  {author} {\bibinfo {author} {\bibfnamefont {D.}~\bibnamefont
  {Poletti}}, \bibinfo {author} {\bibfnamefont {P.}~\bibnamefont {Barmettler}},
  \bibinfo {author} {\bibfnamefont {A.}~\bibnamefont {Georges}},\ and\ \bibinfo
  {author} {\bibfnamefont {C.}~\bibnamefont {Kollath}},\ }\bibfield  {title}
  {\bibinfo {title} {{Emergence of Glasslike Dynamics for Dissipative and
  Strongly Interacting Bosons}},\ }\href
  {https://doi.org/10.1103/PhysRevLett.111.195301} {\bibfield  {journal}
  {\bibinfo  {journal} {Phys. Rev. Lett.}\ }\textbf {\bibinfo {volume} {111}},\
  \bibinfo {pages} {195301} (\bibinfo {year} {2013})}\BibitemShut {NoStop}%
\bibitem [{\citenamefont {Ren}\ \emph {et~al.}(2020)\citenamefont {Ren},
  \citenamefont {Li}, \citenamefont {Li}, \citenamefont {Cai},\ and\
  \citenamefont {Wang}}]{ren2020noise}%
  \BibitemOpen
  \bibfield  {author} {\bibinfo {author} {\bibfnamefont {J.}~\bibnamefont
  {Ren}}, \bibinfo {author} {\bibfnamefont {Q.}~\bibnamefont {Li}}, \bibinfo
  {author} {\bibfnamefont {W.}~\bibnamefont {Li}}, \bibinfo {author}
  {\bibfnamefont {Z.}~\bibnamefont {Cai}},\ and\ \bibinfo {author}
  {\bibfnamefont {X.}~\bibnamefont {Wang}},\ }\bibfield  {title} {\bibinfo
  {title} {{Noise-Driven Universal Dynamics towards an Infinite Temperature
  State}},\ }\href {https://doi.org/10.1103/PhysRevLett.124.130602} {\bibfield
  {journal} {\bibinfo  {journal} {Phys. Rev. Lett.}\ }\textbf {\bibinfo
  {volume} {124}},\ \bibinfo {pages} {130602} (\bibinfo {year}
  {2020})}\BibitemShut {NoStop}%
\bibitem [{\citenamefont {Bouganne}\ \emph {et~al.}(2020)\citenamefont
  {Bouganne}, \citenamefont {Bosch~Aguilera}, \citenamefont {Ghermaoui},
  \citenamefont {Beugnon},\ and\ \citenamefont
  {Gerbier}}]{bouganne2020anomalous}%
  \BibitemOpen
  \bibfield  {author} {\bibinfo {author} {\bibfnamefont {R.}~\bibnamefont
  {Bouganne}}, \bibinfo {author} {\bibfnamefont {M.}~\bibnamefont
  {Bosch~Aguilera}}, \bibinfo {author} {\bibfnamefont {A.}~\bibnamefont
  {Ghermaoui}}, \bibinfo {author} {\bibfnamefont {J.}~\bibnamefont {Beugnon}},\
  and\ \bibinfo {author} {\bibfnamefont {F.}~\bibnamefont {Gerbier}},\
  }\bibfield  {title} {\bibinfo {title} {{Anomalous decay of coherence in a
  dissipative many-body system}},\ }\href
  {https://doi.org/10.1038/s41567-019-0678-2} {\bibfield  {journal} {\bibinfo
  {journal} {Nat. Phys.}\ }\textbf {\bibinfo {volume} {16}},\ \bibinfo {pages}
  {21} (\bibinfo {year} {2020})}\BibitemShut {NoStop}%
\bibitem [{\citenamefont {Plankensteiner}\ \emph {et~al.}(2016)\citenamefont
  {Plankensteiner}, \citenamefont {Schachenmayer}, \citenamefont {Ritsch},\
  and\ \citenamefont {Genes}}]{plankensteiner2016laser}%
  \BibitemOpen
  \bibfield  {author} {\bibinfo {author} {\bibfnamefont {D.}~\bibnamefont
  {Plankensteiner}}, \bibinfo {author} {\bibfnamefont {J.}~\bibnamefont
  {Schachenmayer}}, \bibinfo {author} {\bibfnamefont {H.}~\bibnamefont
  {Ritsch}},\ and\ \bibinfo {author} {\bibfnamefont {C.}~\bibnamefont
  {Genes}},\ }\bibfield  {title} {\bibinfo {title} {{Laser noise imposed
  limitations of ensemble quantum metrology}},\ }\href
  {https://doi.org/10.1088/0953-4075/49/24/245501} {\bibfield  {journal}
  {\bibinfo  {journal} {J. Phys. B: At. Mol. Opt. Phys.}\ }\textbf {\bibinfo
  {volume} {49}},\ \bibinfo {pages} {245501} (\bibinfo {year}
  {2016})}\BibitemShut {NoStop}%
\bibitem [{\citenamefont {Gardiner}\ and\ \citenamefont
  {Zoller}(1991)}]{gardiner2004quantum}%
  \BibitemOpen
  \bibfield  {author} {\bibinfo {author} {\bibfnamefont {C.~W.}\ \bibnamefont
  {Gardiner}}\ and\ \bibinfo {author} {\bibfnamefont {P.}~\bibnamefont
  {Zoller}},\ }\href@noop {} {\emph {\bibinfo {title} {Quantum noise}}}\
  (\bibinfo  {publisher} {Springer},\ \bibinfo {year} {1991})\ Chap.\ \bibinfo
  {chapter} {3.6}, p.~\bibinfo {pages} {77}\BibitemShut {NoStop}%
\bibitem [{\citenamefont {Pichler}\ \emph {et~al.}(2010)\citenamefont
  {Pichler}, \citenamefont {Daley},\ and\ \citenamefont
  {Zoller}}]{pichler2010nonequilibrium}%
  \BibitemOpen
  \bibfield  {author} {\bibinfo {author} {\bibfnamefont {H.}~\bibnamefont
  {Pichler}}, \bibinfo {author} {\bibfnamefont {A.~J.}\ \bibnamefont {Daley}},\
  and\ \bibinfo {author} {\bibfnamefont {P.}~\bibnamefont {Zoller}},\
  }\bibfield  {title} {\bibinfo {title} {{Nonequilibrium dynamics of bosonic
  atoms in optical lattices: Decoherence of many-body states due to spontaneous
  emission}},\ }\href {https://doi.org/10.1103/PhysRevA.82.063605} {\bibfield
  {journal} {\bibinfo  {journal} {Phys. Rev. A}\ }\textbf {\bibinfo {volume}
  {82}},\ \bibinfo {pages} {063605} (\bibinfo {year} {2010})}\BibitemShut
  {NoStop}%
\bibitem [{\citenamefont {Or{\ifmmode\acute{u}\else\'{u}\fi}s}\ and\
  \citenamefont {Vidal}(2008)}]{orus2008infinite}%
  \BibitemOpen
  \bibfield  {author} {\bibinfo {author} {\bibfnamefont {R.}~\bibnamefont
  {Or{\ifmmode\acute{u}\else\'{u}\fi}s}}\ and\ \bibinfo {author} {\bibfnamefont
  {G.}~\bibnamefont {Vidal}},\ }\bibfield  {title} {\bibinfo {title} {{Infinite
  time-evolving block decimation algorithm beyond unitary evolution}},\ }\href
  {https://doi.org/10.1103/PhysRevB.78.155117} {\bibfield  {journal} {\bibinfo
  {journal} {Phys. Rev. B}\ }\textbf {\bibinfo {volume} {78}},\ \bibinfo
  {pages} {155117} (\bibinfo {year} {2008})}\BibitemShut {NoStop}%
\bibitem [{\citenamefont {Harris}(1965)}]{harris1965diffusion}%
  \BibitemOpen
  \bibfield  {author} {\bibinfo {author} {\bibfnamefont {T.~E.}\ \bibnamefont
  {Harris}},\ }\bibfield  {title} {\bibinfo {title} {{Diffusion with
  ``Collisions'' between Particles}},\ }\href
  {https://www.jstor.org/stable/3212197} {\bibfield  {journal} {\bibinfo
  {journal} {J. Appl. Probab.}\ }\textbf {\bibinfo {volume} {2}},\ \bibinfo
  {pages} {323} (\bibinfo {year} {1965})}\BibitemShut {NoStop}%
\bibitem [{\citenamefont {Levitt}(1973)}]{levitt1973dynamics}%
  \BibitemOpen
  \bibfield  {author} {\bibinfo {author} {\bibfnamefont {D.~G.}\ \bibnamefont
  {Levitt}},\ }\bibfield  {title} {\bibinfo {title} {{Dynamics of a Single-File
  Pore: Non-Fickian Behavior}},\ }\href
  {https://doi.org/10.1103/PhysRevA.8.3050} {\bibfield  {journal} {\bibinfo
  {journal} {Phys. Rev. A}\ }\textbf {\bibinfo {volume} {8}},\ \bibinfo {pages}
  {3050} (\bibinfo {year} {1973})}\BibitemShut {NoStop}%
\bibitem [{\citenamefont {Arratia}(1983)}]{arratia1983motion}%
  \BibitemOpen
  \bibfield  {author} {\bibinfo {author} {\bibfnamefont {R.}~\bibnamefont
  {Arratia}},\ }\bibfield  {title} {\bibinfo {title} {{The Motion of a Tagged
  Particle in the Simple Symmetric Exclusion System on $Z$}},\ }\href
  {https://doi.org/10.1214/aop/1176993602} {\bibfield  {journal} {\bibinfo
  {journal} {aop}\ }\textbf {\bibinfo {volume} {11}},\ \bibinfo {pages} {362}
  (\bibinfo {year} {1983})}\BibitemShut {NoStop}%
\bibitem [{\citenamefont {Weimer}\ \emph {et~al.}(2021)\citenamefont {Weimer},
  \citenamefont {Kshetrimayum},\ and\ \citenamefont
  {Or{\ifmmode\acute{u}\else\'{u}\fi}s}}]{weimer2021simulation}%
  \BibitemOpen
  \bibfield  {author} {\bibinfo {author} {\bibfnamefont {H.}~\bibnamefont
  {Weimer}}, \bibinfo {author} {\bibfnamefont {A.}~\bibnamefont
  {Kshetrimayum}},\ and\ \bibinfo {author} {\bibfnamefont {R.}~\bibnamefont
  {Or{\ifmmode\acute{u}\else\'{u}\fi}s}},\ }\bibfield  {title} {\bibinfo
  {title} {{Simulation methods for open quantum many-body systems}},\ }\href
  {https://doi.org/10.1103/RevModPhys.93.015008} {\bibfield  {journal}
  {\bibinfo  {journal} {Rev. Mod. Phys.}\ }\textbf {\bibinfo {volume} {93}},\
  \bibinfo {pages} {015008} (\bibinfo {year} {2021})}\BibitemShut {NoStop}%
\bibitem [{\citenamefont {Daley}(2014)}]{daley2014quantum}%
  \BibitemOpen
  \bibfield  {author} {\bibinfo {author} {\bibfnamefont {A.~J.}\ \bibnamefont
  {Daley}},\ }\bibfield  {title} {\bibinfo {title} {{Quantum trajectories and
  open many-body quantum systems}},\ }\href
  {https://doi.org/10.1080/00018732.2014.933502} {\bibfield  {journal}
  {\bibinfo  {journal} {Adv. Phys.}\ }\textbf {\bibinfo {volume} {63}},\
  \bibinfo {pages} {77} (\bibinfo {year} {2014})}\BibitemShut {NoStop}%
\bibitem [{\citenamefont {Zwolak}\ and\ \citenamefont
  {Vidal}(2004)}]{zwolak2004mixed}%
  \BibitemOpen
  \bibfield  {author} {\bibinfo {author} {\bibfnamefont {M.}~\bibnamefont
  {Zwolak}}\ and\ \bibinfo {author} {\bibfnamefont {G.}~\bibnamefont {Vidal}},\
  }\bibfield  {title} {\bibinfo {title} {{Mixed-State Dynamics in
  One-Dimensional Quantum Lattice Systems: A Time-Dependent Superoperator
  Renormalization Algorithm}},\ }\href
  {https://doi.org/10.1103/PhysRevLett.93.207205} {\bibfield  {journal}
  {\bibinfo  {journal} {Phys. Rev. Lett.}\ }\textbf {\bibinfo {volume} {93}},\
  \bibinfo {pages} {207205} (\bibinfo {year} {2004})}\BibitemShut {NoStop}%
\bibitem [{\citenamefont {Verstraete}\ \emph {et~al.}(2004)\citenamefont
  {Verstraete}, \citenamefont
  {Garc{\ifmmode\acute{\imath}\else\'{\i}\fi}a-Ripoll},\ and\ \citenamefont
  {Cirac}}]{verstraete2004matrix}%
  \BibitemOpen
  \bibfield  {author} {\bibinfo {author} {\bibfnamefont {F.}~\bibnamefont
  {Verstraete}}, \bibinfo {author} {\bibfnamefont {J.~J.}\ \bibnamefont
  {Garc{\ifmmode\acute{\imath}\else\'{\i}\fi}a-Ripoll}},\ and\ \bibinfo
  {author} {\bibfnamefont {J.~I.}\ \bibnamefont {Cirac}},\ }\bibfield  {title}
  {\bibinfo {title} {{Matrix Product Density Operators: Simulation of
  Finite-Temperature and Dissipative Systems}},\ }\href
  {https://doi.org/10.1103/PhysRevLett.93.207204} {\bibfield  {journal}
  {\bibinfo  {journal} {Phys. Rev. Lett.}\ }\textbf {\bibinfo {volume} {93}},\
  \bibinfo {pages} {207204} (\bibinfo {year} {2004})}\BibitemShut {NoStop}%
\bibitem [{\citenamefont {Werner}\ \emph {et~al.}(2016)\citenamefont {Werner},
  \citenamefont {Jaschke}, \citenamefont {Silvi}, \citenamefont {Kliesch},
  \citenamefont {Calarco}, \citenamefont {Eisert},\ and\ \citenamefont
  {Montangero}}]{werner2016positive}%
  \BibitemOpen
  \bibfield  {author} {\bibinfo {author} {\bibfnamefont {A.~H.}\ \bibnamefont
  {Werner}}, \bibinfo {author} {\bibfnamefont {D.}~\bibnamefont {Jaschke}},
  \bibinfo {author} {\bibfnamefont {P.}~\bibnamefont {Silvi}}, \bibinfo
  {author} {\bibfnamefont {M.}~\bibnamefont {Kliesch}}, \bibinfo {author}
  {\bibfnamefont {T.}~\bibnamefont {Calarco}}, \bibinfo {author} {\bibfnamefont
  {J.}~\bibnamefont {Eisert}},\ and\ \bibinfo {author} {\bibfnamefont
  {S.}~\bibnamefont {Montangero}},\ }\bibfield  {title} {\bibinfo {title}
  {{Positive Tensor Network Approach for Simulating Open Quantum Many-Body
  Systems}},\ }\href {https://doi.org/10.1103/PhysRevLett.116.237201}
  {\bibfield  {journal} {\bibinfo  {journal} {Phys. Rev. Lett.}\ }\textbf
  {\bibinfo {volume} {116}},\ \bibinfo {pages} {237201} (\bibinfo {year}
  {2016})}\BibitemShut {NoStop}%
\bibitem [{\citenamefont {Vidal}(2007)}]{vidal2007classical}%
  \BibitemOpen
  \bibfield  {author} {\bibinfo {author} {\bibfnamefont {G.}~\bibnamefont
  {Vidal}},\ }\bibfield  {title} {\bibinfo {title} {{Classical Simulation of
  Infinite-Size Quantum Lattice Systems in One Spatial Dimension}},\ }\href
  {https://doi.org/10.1103/PhysRevLett.98.070201} {\bibfield  {journal}
  {\bibinfo  {journal} {Phys. Rev. Lett.}\ }\textbf {\bibinfo {volume} {98}},\
  \bibinfo {pages} {070201} (\bibinfo {year} {2007})}\BibitemShut {NoStop}%
\bibitem [{\citenamefont {Sornborger}\ and\ \citenamefont
  {Stewart}(1999)}]{Sornborger_Higher_1999}%
  \BibitemOpen
  \bibfield  {author} {\bibinfo {author} {\bibfnamefont {A.~T.}\ \bibnamefont
  {Sornborger}}\ and\ \bibinfo {author} {\bibfnamefont {E.~D.}\ \bibnamefont
  {Stewart}},\ }\bibfield  {title} {\bibinfo {title} {{Higher-order methods for
  simulations on quantum computers}},\ }\href
  {https://doi.org/10.1103/PhysRevA.60.1956} {\bibfield  {journal} {\bibinfo
  {journal} {Phys. Rev. A}\ }\textbf {\bibinfo {volume} {60}},\ \bibinfo
  {pages} {1956} (\bibinfo {year} {1999})}\BibitemShut {NoStop}%
\bibitem [{\citenamefont {Carollo}\ and\ \citenamefont
  {Alba}(2021)}]{carollo2021emergent}%
  \BibitemOpen
  \bibfield  {author} {\bibinfo {author} {\bibfnamefont {F.}~\bibnamefont
  {Carollo}}\ and\ \bibinfo {author} {\bibfnamefont {V.}~\bibnamefont {Alba}},\
  }\bibfield  {title} {\bibinfo {title} {{Emergent dissipative quasi-particle
  picture in noninteracting Markovian open quantum systems}},\ }\href
  {https://arxiv.org/abs/2106.11997v1} {\bibfield  {journal} {\bibinfo
  {journal} {arXiv:2106.11997}\ } (\bibinfo {year} {2021})}\BibitemShut
  {NoStop}%
\bibitem [{\citenamefont {Alba}\ and\ \citenamefont
  {Carollo}(2021)}]{alba2021hydrodynamics}%
  \BibitemOpen
  \bibfield  {author} {\bibinfo {author} {\bibfnamefont {V.}~\bibnamefont
  {Alba}}\ and\ \bibinfo {author} {\bibfnamefont {F.}~\bibnamefont {Carollo}},\
  }\bibfield  {title} {\bibinfo {title} {{Hydrodynamics of quantum entropies in
  Ising chains with linear dissipation}},\ }\href
  {https://arxiv.org/abs/2109.01836v1} {\bibfield  {journal} {\bibinfo
  {journal} {arXiv:2109.01836}\ } (\bibinfo {year} {2021})}\BibitemShut
  {NoStop}%
\bibitem [{\citenamefont {Medvedyeva}\ \emph
  {et~al.}(2016{\natexlab{b}})\citenamefont {Medvedyeva}, \citenamefont
  {Prosen},\ and\ \citenamefont
  {{\ifmmode\check{Z}\else\v{Z}\fi}nidari{\ifmmode\check{c}\else\v{c}\fi}}}]{medvedyeva2016influence}%
  \BibitemOpen
  \bibfield  {author} {\bibinfo {author} {\bibfnamefont {M.~V.}\ \bibnamefont
  {Medvedyeva}}, \bibinfo {author} {\bibfnamefont {T.}~\bibnamefont {Prosen}},\
  and\ \bibinfo {author} {\bibfnamefont {M.}~\bibnamefont
  {{\ifmmode\check{Z}\else\v{Z}\fi}nidari{\ifmmode\check{c}\else\v{c}\fi}}},\
  }\bibfield  {title} {\bibinfo {title} {{Influence of dephasing on many-body
  localization}},\ }\href {https://doi.org/10.1103/PhysRevB.93.094205}
  {\bibfield  {journal} {\bibinfo  {journal} {Phys. Rev. B}\ }\textbf {\bibinfo
  {volume} {93}},\ \bibinfo {pages} {094205} (\bibinfo {year}
  {2016}{\natexlab{b}})}\BibitemShut {NoStop}%
\bibitem [{\citenamefont {Mallick}(2015)}]{mallick2015exclusion}%
  \BibitemOpen
  \bibfield  {author} {\bibinfo {author} {\bibfnamefont {K.}~\bibnamefont
  {Mallick}},\ }\bibfield  {title} {\bibinfo {title} {{The exclusion process: A
  paradigm for non-equilibrium behaviour}},\ }\href
  {https://doi.org/10.1016/j.physa.2014.07.046} {\bibfield  {journal} {\bibinfo
   {journal} {Physica A}\ }\textbf {\bibinfo {volume} {418}},\ \bibinfo {pages}
  {17} (\bibinfo {year} {2015})}\BibitemShut {NoStop}%
\bibitem [{\citenamefont {Bernard}\ \emph {et~al.}(2018)\citenamefont
  {Bernard}, \citenamefont {Jin},\ and\ \citenamefont
  {Shpielberg}}]{bernard2018transport}%
  \BibitemOpen
  \bibfield  {author} {\bibinfo {author} {\bibfnamefont {D.}~\bibnamefont
  {Bernard}}, \bibinfo {author} {\bibfnamefont {T.}~\bibnamefont {Jin}},\ and\
  \bibinfo {author} {\bibfnamefont {O.}~\bibnamefont {Shpielberg}},\ }\bibfield
   {title} {\bibinfo {title} {Transport in quantum chains under strong
  monitoring},\ }\href {https://doi.org/10.1209/0295-5075/121/60006} {\bibfield
   {journal} {\bibinfo  {journal} {EPL (Europhysics Letters)}\ }\textbf
  {\bibinfo {volume} {121}},\ \bibinfo {pages} {60006} (\bibinfo {year}
  {2018})}\BibitemShut {NoStop}%
\bibitem [{\citenamefont {Lin}\ \emph {et~al.}(2005)\citenamefont {Lin},
  \citenamefont {Meron}, \citenamefont {Cui}, \citenamefont {Rice},\ and\
  \citenamefont {Diamant}}]{lin2005random}%
  \BibitemOpen
  \bibfield  {author} {\bibinfo {author} {\bibfnamefont {B.}~\bibnamefont
  {Lin}}, \bibinfo {author} {\bibfnamefont {M.}~\bibnamefont {Meron}}, \bibinfo
  {author} {\bibfnamefont {B.}~\bibnamefont {Cui}}, \bibinfo {author}
  {\bibfnamefont {S.~A.}\ \bibnamefont {Rice}},\ and\ \bibinfo {author}
  {\bibfnamefont {H.}~\bibnamefont {Diamant}},\ }\bibfield  {title} {\bibinfo
  {title} {{From Random Walk to Single-File Diffusion}},\ }\href
  {https://doi.org/10.1103/PhysRevLett.94.216001} {\bibfield  {journal}
  {\bibinfo  {journal} {Phys. Rev. Lett.}\ }\textbf {\bibinfo {volume} {94}},\
  \bibinfo {pages} {216001} (\bibinfo {year} {2005})}\BibitemShut {NoStop}%
\bibitem [{\citenamefont {Derrida}\ and\ \citenamefont
  {Gerschenfeld}(2009)}]{derrida2009current}%
  \BibitemOpen
  \bibfield  {author} {\bibinfo {author} {\bibfnamefont {B.}~\bibnamefont
  {Derrida}}\ and\ \bibinfo {author} {\bibfnamefont {A.}~\bibnamefont
  {Gerschenfeld}},\ }\bibfield  {title} {\bibinfo {title} {{Current
  Fluctuations of the One Dimensional Symmetric Simple Exclusion Process with
  Step Initial Condition}},\ }\href {https://doi.org/10.1007/s10955-009-9772-7}
  {\bibfield  {journal} {\bibinfo  {journal} {J. Stat. Phys.}\ }\textbf
  {\bibinfo {volume} {136}},\ \bibinfo {pages} {1} (\bibinfo {year}
  {2009})}\BibitemShut {NoStop}%
\bibitem [{\citenamefont {Imamura}\ \emph {et~al.}(2017)\citenamefont
  {Imamura}, \citenamefont {Mallick},\ and\ \citenamefont
  {Sasamoto}}]{imamura2017large}%
  \BibitemOpen
  \bibfield  {author} {\bibinfo {author} {\bibfnamefont {T.}~\bibnamefont
  {Imamura}}, \bibinfo {author} {\bibfnamefont {K.}~\bibnamefont {Mallick}},\
  and\ \bibinfo {author} {\bibfnamefont {T.}~\bibnamefont {Sasamoto}},\
  }\bibfield  {title} {\bibinfo {title} {{Large Deviations of a Tracer in the
  Symmetric Exclusion Process}},\ }\href
  {https://doi.org/10.1103/PhysRevLett.118.160601} {\bibfield  {journal}
  {\bibinfo  {journal} {Phys. Rev. Lett.}\ }\textbf {\bibinfo {volume} {118}},\
  \bibinfo {pages} {160601} (\bibinfo {year} {2017})}\BibitemShut {NoStop}%
\bibitem [{\citenamefont {Grabsch}\ \emph {et~al.}(2021)\citenamefont
  {Grabsch}, \citenamefont {Poncet}, \citenamefont {Rizkallah}, \citenamefont
  {Illien},\ and\ \citenamefont {B{\'e}nichou}}]{grabsch2021closing}%
  \BibitemOpen
  \bibfield  {author} {\bibinfo {author} {\bibfnamefont {A.}~\bibnamefont
  {Grabsch}}, \bibinfo {author} {\bibfnamefont {A.}~\bibnamefont {Poncet}},
  \bibinfo {author} {\bibfnamefont {P.}~\bibnamefont {Rizkallah}}, \bibinfo
  {author} {\bibfnamefont {P.}~\bibnamefont {Illien}},\ and\ \bibinfo {author}
  {\bibfnamefont {O.}~\bibnamefont {B{\'e}nichou}},\ }\bibfield  {title}
  {\bibinfo {title} {Closing and solving the hierarchy for large deviations and
  spatial correlations in single-file diffusion},\ }\href
  {https://arxiv.org/abs/2110.09269} {\bibfield  {journal} {\bibinfo  {journal}
  {arXiv:2110.09269}\ } (\bibinfo {year} {2021})}\BibitemShut {NoStop}%
\bibitem [{\citenamefont {Hahn}\ \emph {et~al.}(1996)\citenamefont {Hahn},
  \citenamefont {K{\ifmmode\ddot{a}\else\"{a}\fi}rger},\ and\ \citenamefont
  {Kukla}}]{hahn1996single}%
  \BibitemOpen
  \bibfield  {author} {\bibinfo {author} {\bibfnamefont {K.}~\bibnamefont
  {Hahn}}, \bibinfo {author} {\bibfnamefont {J.}~\bibnamefont
  {K{\ifmmode\ddot{a}\else\"{a}\fi}rger}},\ and\ \bibinfo {author}
  {\bibfnamefont {V.}~\bibnamefont {Kukla}},\ }\bibfield  {title} {\bibinfo
  {title} {{Single-File Diffusion Observation}},\ }\href
  {https://doi.org/10.1103/PhysRevLett.76.2762} {\bibfield  {journal} {\bibinfo
   {journal} {Phys. Rev. Lett.}\ }\textbf {\bibinfo {volume} {76}},\ \bibinfo
  {pages} {2762} (\bibinfo {year} {1996})}\BibitemShut {NoStop}%
\bibitem [{\citenamefont {Wei}\ \emph {et~al.}(2000)\citenamefont {Wei},
  \citenamefont {Bechinger},\ and\ \citenamefont {Leiderer}}]{wei2000single}%
  \BibitemOpen
  \bibfield  {author} {\bibinfo {author} {\bibfnamefont {Q.-H.}\ \bibnamefont
  {Wei}}, \bibinfo {author} {\bibfnamefont {C.}~\bibnamefont {Bechinger}},\
  and\ \bibinfo {author} {\bibfnamefont {P.}~\bibnamefont {Leiderer}},\
  }\bibfield  {title} {\bibinfo {title} {{Single-File Diffusion of Colloids in
  One-Dimensional Channels}},\ }\href
  {https://doi.org/10.1126/science.287.5453.625} {\bibfield  {journal}
  {\bibinfo  {journal} {Science}\ }\textbf {\bibinfo {volume} {287}},\ \bibinfo
  {pages} {625} (\bibinfo {year} {2000})}\BibitemShut {NoStop}%
\bibitem [{\citenamefont
  {{\ifmmode\check{Z}\else\v{Z}\fi}nidari{\ifmmode\check{c}\else\v{c}\fi}}(2010{\natexlab{a}})}]{znidaric2010dephasing}%
  \BibitemOpen
  \bibfield  {author} {\bibinfo {author} {\bibfnamefont {M.}~\bibnamefont
  {{\ifmmode\check{Z}\else\v{Z}\fi}nidari{\ifmmode\check{c}\else\v{c}\fi}}},\
  }\bibfield  {title} {\bibinfo {title} {{Dephasing-induced diffusive transport
  in the anisotropic Heisenberg model}},\ }\href
  {https://doi.org/10.1088/1367-2630/12/4/043001} {\bibfield  {journal}
  {\bibinfo  {journal} {New J. Phys.}\ }\textbf {\bibinfo {volume} {12}},\
  \bibinfo {pages} {043001} (\bibinfo {year} {2010}{\natexlab{a}})}\BibitemShut
  {NoStop}%
\bibitem [{\citenamefont
  {{\ifmmode\check{Z}\else\v{Z}\fi}nidari{\ifmmode\check{c}\else\v{c}\fi}}(2010{\natexlab{b}})}]{znidaric2010exact}%
  \BibitemOpen
  \bibfield  {author} {\bibinfo {author} {\bibfnamefont {M.}~\bibnamefont
  {{\ifmmode\check{Z}\else\v{Z}\fi}nidari{\ifmmode\check{c}\else\v{c}\fi}}},\
  }\bibfield  {title} {\bibinfo {title} {{Exact solution for a diffusive
  nonequilibrium steady state of an open quantum}},\ }\href
  {https://doi.org/10.1088/1742-5468/2010/05/l05002} {\bibfield  {journal}
  {\bibinfo  {journal} {J. Stat. Mech.: Theory Exp.}\ }\textbf {\bibinfo
  {volume} {2010}}\bibinfo  {number} { (05)},\ \bibinfo {pages}
  {L05002}}\BibitemShut {NoStop}%
\bibitem [{\citenamefont {Eisler}(2011)}]{eisler2011crossover}%
  \BibitemOpen
\bibfield  {number} {  }\bibfield  {author} {\bibinfo {author} {\bibfnamefont
  {V.}~\bibnamefont {Eisler}},\ }\bibfield  {title} {\bibinfo {title}
  {{Crossover between ballistic and diffusive transport: the quantum exclusion
  process}},\ }\href {https://doi.org/10.1088/1742-5468/2011/06/p06007}
  {\bibfield  {journal} {\bibinfo  {journal} {J. Stat. Mech.: Theory Exp.}\
  }\textbf {\bibinfo {volume} {2011}}\bibinfo  {number} { (06)},\ \bibinfo
  {pages} {P06007}}\BibitemShut {NoStop}%
\bibitem [{\citenamefont {De~Nardis}\ \emph {et~al.}(2021)\citenamefont
  {De~Nardis}, \citenamefont {Gopalakrishnan}, \citenamefont {Vasseur},\ and\
  \citenamefont {Ware}}]{denardis2021subdiffusive}%
  \BibitemOpen
\bibfield  {number} {  }\bibfield  {author} {\bibinfo {author} {\bibfnamefont
  {J.}~\bibnamefont {De~Nardis}}, \bibinfo {author} {\bibfnamefont
  {S.}~\bibnamefont {Gopalakrishnan}}, \bibinfo {author} {\bibfnamefont
  {R.}~\bibnamefont {Vasseur}},\ and\ \bibinfo {author} {\bibfnamefont
  {B.}~\bibnamefont {Ware}},\ }\bibfield  {title} {\bibinfo {title}
  {{Subdiffusive hydrodynamics of nearly-integrable anisotropic spin chains}},\
  }\href {https://arxiv.org/abs/2109.13251v2} {\bibfield  {journal} {\bibinfo
  {journal} {arXiv:2109.13251}\ } (\bibinfo {year} {2021})}\BibitemShut
  {NoStop}%
\bibitem [{\citenamefont {Goldstein}\ and\ \citenamefont
  {Sela}(2018)}]{goldstein2018symmetry}%
  \BibitemOpen
  \bibfield  {author} {\bibinfo {author} {\bibfnamefont {M.}~\bibnamefont
  {Goldstein}}\ and\ \bibinfo {author} {\bibfnamefont {E.}~\bibnamefont
  {Sela}},\ }\bibfield  {title} {\bibinfo {title} {Symmetry-resolved
  entanglement in many-body systems},\ }\href
  {https://doi.org/10.1103/PhysRevLett.120.200602} {\bibfield  {journal}
  {\bibinfo  {journal} {Phys. Rev. Lett.}\ }\textbf {\bibinfo {volume} {120}},\
  \bibinfo {pages} {200602} (\bibinfo {year} {2018})}\BibitemShut {NoStop}%
\bibitem [{\citenamefont {Xavier}\ \emph {et~al.}(2018)\citenamefont {Xavier},
  \citenamefont {Alcaraz},\ and\ \citenamefont
  {Sierra}}]{xavier2018equipartition}%
  \BibitemOpen
  \bibfield  {author} {\bibinfo {author} {\bibfnamefont {J.~C.}\ \bibnamefont
  {Xavier}}, \bibinfo {author} {\bibfnamefont {F.~C.}\ \bibnamefont
  {Alcaraz}},\ and\ \bibinfo {author} {\bibfnamefont {G.}~\bibnamefont
  {Sierra}},\ }\bibfield  {title} {\bibinfo {title} {Equipartition of the
  entanglement entropy},\ }\href {https://doi.org/10.1103/PhysRevB.98.041106}
  {\bibfield  {journal} {\bibinfo  {journal} {Phys. Rev. B}\ }\textbf {\bibinfo
  {volume} {98}},\ \bibinfo {pages} {041106(R)} (\bibinfo {year}
  {2018})}\BibitemShut {NoStop}%
\bibitem [{\citenamefont {Parez}\ \emph {et~al.}(2021)\citenamefont {Parez},
  \citenamefont {Bonsignori},\ and\ \citenamefont
  {Calabrese}}]{parez2021quasiparticle}%
  \BibitemOpen
  \bibfield  {author} {\bibinfo {author} {\bibfnamefont {G.}~\bibnamefont
  {Parez}}, \bibinfo {author} {\bibfnamefont {R.}~\bibnamefont {Bonsignori}},\
  and\ \bibinfo {author} {\bibfnamefont {P.}~\bibnamefont {Calabrese}},\
  }\bibfield  {title} {\bibinfo {title} {Quasiparticle dynamics of
  symmetry-resolved entanglement after a quench: Examples of conformal field
  theories and free fermions},\ }\href
  {https://doi.org/10.1103/PhysRevB.103.L041104} {\bibfield  {journal}
  {\bibinfo  {journal} {Phys. Rev. B}\ }\textbf {\bibinfo {volume} {103}},\
  \bibinfo {pages} {L041104} (\bibinfo {year} {2021})}\BibitemShut {NoStop}%
\bibitem [{\citenamefont {Barghathi}\ \emph {et~al.}(2018)\citenamefont
  {Barghathi}, \citenamefont {Herdman},\ and\ \citenamefont
  {Del~Maestro}}]{barghathi2018renyi}%
  \BibitemOpen
  \bibfield  {author} {\bibinfo {author} {\bibfnamefont {H.}~\bibnamefont
  {Barghathi}}, \bibinfo {author} {\bibfnamefont {C.~M.}\ \bibnamefont
  {Herdman}},\ and\ \bibinfo {author} {\bibfnamefont {A.}~\bibnamefont
  {Del~Maestro}},\ }\bibfield  {title} {\bibinfo {title} {{R\'enyi
  Generalization of the Accessible Entanglement Entropy}},\ }\href
  {https://doi.org/10.1103/PhysRevLett.121.150501} {\bibfield  {journal}
  {\bibinfo  {journal} {Phys. Rev. Lett.}\ }\textbf {\bibinfo {volume} {121}},\
  \bibinfo {pages} {150501} (\bibinfo {year} {2018})}\BibitemShut {NoStop}%
\bibitem [{\citenamefont {Barghathi}\ \emph {et~al.}(2019)\citenamefont
  {Barghathi}, \citenamefont {Casiano-Diaz},\ and\ \citenamefont
  {Del~Maestro}}]{barghathi2019operationally}%
  \BibitemOpen
  \bibfield  {author} {\bibinfo {author} {\bibfnamefont {H.}~\bibnamefont
  {Barghathi}}, \bibinfo {author} {\bibfnamefont {E.}~\bibnamefont
  {Casiano-Diaz}},\ and\ \bibinfo {author} {\bibfnamefont {A.}~\bibnamefont
  {Del~Maestro}},\ }\bibfield  {title} {\bibinfo {title} {{Operationally
  accessible entanglement of one-dimensional spinless fermions}},\ }\href
  {https://doi.org/10.1103/PhysRevA.100.022324} {\bibfield  {journal} {\bibinfo
   {journal} {Phys. Rev. A}\ }\textbf {\bibinfo {volume} {100}},\ \bibinfo
  {pages} {022324} (\bibinfo {year} {2019})}\BibitemShut {NoStop}%
\bibitem [{\citenamefont {Li}\ \emph {et~al.}(2018)\citenamefont {Li},
  \citenamefont {Chen},\ and\ \citenamefont {Fisher}}]{li2018quantum}%
  \BibitemOpen
  \bibfield  {author} {\bibinfo {author} {\bibfnamefont {Y.}~\bibnamefont
  {Li}}, \bibinfo {author} {\bibfnamefont {X.}~\bibnamefont {Chen}},\ and\
  \bibinfo {author} {\bibfnamefont {M.~P.~A.}\ \bibnamefont {Fisher}},\
  }\bibfield  {title} {\bibinfo {title} {{Quantum Zeno effect and the many-body
  entanglement transition}},\ }\href
  {https://doi.org/10.1103/PhysRevB.98.205136} {\bibfield  {journal} {\bibinfo
  {journal} {Phys. Rev. B}\ }\textbf {\bibinfo {volume} {98}},\ \bibinfo
  {pages} {205136} (\bibinfo {year} {2018})}\BibitemShut {NoStop}%
\bibitem [{\citenamefont {Chan}\ \emph {et~al.}(2019)\citenamefont {Chan},
  \citenamefont {Nandkishore}, \citenamefont {Pretko},\ and\ \citenamefont
  {Smith}}]{chan2019unitary}%
  \BibitemOpen
  \bibfield  {author} {\bibinfo {author} {\bibfnamefont {A.}~\bibnamefont
  {Chan}}, \bibinfo {author} {\bibfnamefont {R.~M.}\ \bibnamefont
  {Nandkishore}}, \bibinfo {author} {\bibfnamefont {M.}~\bibnamefont
  {Pretko}},\ and\ \bibinfo {author} {\bibfnamefont {G.}~\bibnamefont
  {Smith}},\ }\bibfield  {title} {\bibinfo {title} {{Unitary-projective
  entanglement dynamics}},\ }\href {https://doi.org/10.1103/PhysRevB.99.224307}
  {\bibfield  {journal} {\bibinfo  {journal} {Phys. Rev. B}\ }\textbf {\bibinfo
  {volume} {99}},\ \bibinfo {pages} {224307} (\bibinfo {year}
  {2019})}\BibitemShut {NoStop}%
\bibitem [{\citenamefont {Coppola}\ \emph {et~al.}(2021)\citenamefont
  {Coppola}, \citenamefont {Tirrito}, \citenamefont {Karevski},\ and\
  \citenamefont {Collura}}]{coppola2021growth}%
  \BibitemOpen
  \bibfield  {author} {\bibinfo {author} {\bibfnamefont {M.}~\bibnamefont
  {Coppola}}, \bibinfo {author} {\bibfnamefont {E.}~\bibnamefont {Tirrito}},
  \bibinfo {author} {\bibfnamefont {D.}~\bibnamefont {Karevski}},\ and\
  \bibinfo {author} {\bibfnamefont {M.}~\bibnamefont {Collura}},\ }\bibfield
  {title} {\bibinfo {title} {{Growth of entanglement entropy under local
  projective measurements}},\ }\href {https://arxiv.org/abs/2109.10837v1}
  {\bibfield  {journal} {\bibinfo  {journal} {arXiv:2109.10837}\ } (\bibinfo
  {year} {2021})}\BibitemShut {NoStop}%
\bibitem [{\citenamefont {Botzung}\ \emph {et~al.}(2021)\citenamefont
  {Botzung}, \citenamefont {Diehl},\ and\ \citenamefont
  {M{\ifmmode\ddot{u}\else\"{u}\fi}ller}}]{botzung2021engineered}%
  \BibitemOpen
  \bibfield  {author} {\bibinfo {author} {\bibfnamefont {T.}~\bibnamefont
  {Botzung}}, \bibinfo {author} {\bibfnamefont {S.}~\bibnamefont {Diehl}},\
  and\ \bibinfo {author} {\bibfnamefont {M.}~\bibnamefont
  {M{\ifmmode\ddot{u}\else\"{u}\fi}ller}},\ }\bibfield  {title} {\bibinfo
  {title} {{Engineered dissipation induced entanglement transition in quantum
  spin chains: From logarithmic growth to area law}},\ }\href
  {https://doi.org/10.1103/PhysRevB.104.184422} {\bibfield  {journal} {\bibinfo
   {journal} {Phys. Rev. B}\ }\textbf {\bibinfo {volume} {104}},\ \bibinfo
  {pages} {184422} (\bibinfo {year} {2021})}\BibitemShut {NoStop}%
\bibitem [{\citenamefont {Mark}\ \emph {et~al.}(2020)\citenamefont {Mark},
  \citenamefont {Flannigan}, \citenamefont {Meinert}, \citenamefont {D'Incao},
  \citenamefont {Daley},\ and\ \citenamefont
  {N{\ifmmode\ddot{a}\else\"{a}\fi}gerl}}]{mark2020interplay}%
  \BibitemOpen
  \bibfield  {author} {\bibinfo {author} {\bibfnamefont {M.~J.}\ \bibnamefont
  {Mark}}, \bibinfo {author} {\bibfnamefont {S.}~\bibnamefont {Flannigan}},
  \bibinfo {author} {\bibfnamefont {F.}~\bibnamefont {Meinert}}, \bibinfo
  {author} {\bibfnamefont {J.~P.}\ \bibnamefont {D'Incao}}, \bibinfo {author}
  {\bibfnamefont {A.~J.}\ \bibnamefont {Daley}},\ and\ \bibinfo {author}
  {\bibfnamefont {H.-C.}\ \bibnamefont
  {N{\ifmmode\ddot{a}\else\"{a}\fi}gerl}},\ }\bibfield  {title} {\bibinfo
  {title} {{Interplay between coherent and dissipative dynamics of bosonic
  doublons in an optical lattice}},\ }\href
  {https://doi.org/10.1103/PhysRevResearch.2.043050} {\bibfield  {journal}
  {\bibinfo  {journal} {Phys. Rev. Res.}\ }\textbf {\bibinfo {volume} {2}},\
  \bibinfo {pages} {043050} (\bibinfo {year} {2020})}\BibitemShut {NoStop}%
\bibitem [{\citenamefont {Zhu}\ \emph {et~al.}(2014)\citenamefont {Zhu},
  \citenamefont {Gadway}, \citenamefont {Foss-Feig}, \citenamefont
  {Schachenmayer}, \citenamefont {Wall}, \citenamefont {Hazzard}, \citenamefont
  {Yan}, \citenamefont {Moses}, \citenamefont {Covey}, \citenamefont {Jin},
  \citenamefont {Ye}, \citenamefont {Holland},\ and\ \citenamefont
  {Rey}}]{zhu2014suppressing}%
  \BibitemOpen
  \bibfield  {author} {\bibinfo {author} {\bibfnamefont {B.}~\bibnamefont
  {Zhu}}, \bibinfo {author} {\bibfnamefont {B.}~\bibnamefont {Gadway}},
  \bibinfo {author} {\bibfnamefont {M.}~\bibnamefont {Foss-Feig}}, \bibinfo
  {author} {\bibfnamefont {J.}~\bibnamefont {Schachenmayer}}, \bibinfo {author}
  {\bibfnamefont {M.~L.}\ \bibnamefont {Wall}}, \bibinfo {author}
  {\bibfnamefont {K.~R.~A.}\ \bibnamefont {Hazzard}}, \bibinfo {author}
  {\bibfnamefont {B.}~\bibnamefont {Yan}}, \bibinfo {author} {\bibfnamefont
  {S.~A.}\ \bibnamefont {Moses}}, \bibinfo {author} {\bibfnamefont {J.~P.}\
  \bibnamefont {Covey}}, \bibinfo {author} {\bibfnamefont {D.~S.}\ \bibnamefont
  {Jin}}, \bibinfo {author} {\bibfnamefont {J.}~\bibnamefont {Ye}}, \bibinfo
  {author} {\bibfnamefont {M.}~\bibnamefont {Holland}},\ and\ \bibinfo {author}
  {\bibfnamefont {A.~M.}\ \bibnamefont {Rey}},\ }\bibfield  {title} {\bibinfo
  {title} {{Suppressing the Loss of Ultracold Molecules Via the Continuous
  Quantum Zeno Effect}},\ }\href
  {https://doi.org/10.1103/PhysRevLett.112.070404} {\bibfield  {journal}
  {\bibinfo  {journal} {Phys. Rev. Lett.}\ }\textbf {\bibinfo {volume} {112}},\
  \bibinfo {pages} {070404} (\bibinfo {year} {2014})}\BibitemShut {NoStop}%
\bibitem [{\citenamefont {Shchesnovich}\ and\ \citenamefont
  {Konotop}(2010)}]{shchesnovich2010control}%
  \BibitemOpen
  \bibfield  {author} {\bibinfo {author} {\bibfnamefont {V.~S.}\ \bibnamefont
  {Shchesnovich}}\ and\ \bibinfo {author} {\bibfnamefont {V.~V.}\ \bibnamefont
  {Konotop}},\ }\bibfield  {title} {\bibinfo {title} {{Control of a
  Bose-Einstein condensate by dissipation: Nonlinear Zeno effect}},\ }\href
  {https://doi.org/10.1103/PhysRevA.81.053611} {\bibfield  {journal} {\bibinfo
  {journal} {Phys. Rev. A}\ }\textbf {\bibinfo {volume} {81}},\ \bibinfo
  {pages} {053611} (\bibinfo {year} {2010})}\BibitemShut {NoStop}%
\bibitem [{\citenamefont {M{\ifmmode\ddot{u}\else\"{u}\fi}ller}\ \emph
  {et~al.}(2022)\citenamefont {M{\ifmmode\ddot{u}\else\"{u}\fi}ller},
  \citenamefont {Diehl},\ and\ \citenamefont
  {Buchhold}}]{muller2022measurement}%
  \BibitemOpen
  \bibfield  {author} {\bibinfo {author} {\bibfnamefont {T.}~\bibnamefont
  {M{\ifmmode\ddot{u}\else\"{u}\fi}ller}}, \bibinfo {author} {\bibfnamefont
  {S.}~\bibnamefont {Diehl}},\ and\ \bibinfo {author} {\bibfnamefont
  {M.}~\bibnamefont {Buchhold}},\ }\bibfield  {title} {\bibinfo {title}
  {{Measurement-Induced Dark State Phase Transitions in Long-Ranged Fermion
  Systems}},\ }\href {https://doi.org/10.1103/PhysRevLett.128.010605}
  {\bibfield  {journal} {\bibinfo  {journal} {Phys. Rev. Lett.}\ }\textbf
  {\bibinfo {volume} {128}},\ \bibinfo {pages} {010605} (\bibinfo {year}
  {2022})}\BibitemShut {NoStop}%
\bibitem [{\citenamefont {Fishman}\ \emph {et~al.}(2020)\citenamefont
  {Fishman}, \citenamefont {White},\ and\ \citenamefont
  {Stoudenmire}}]{itensor}%
  \BibitemOpen
  \bibfield  {author} {\bibinfo {author} {\bibfnamefont {M.}~\bibnamefont
  {Fishman}}, \bibinfo {author} {\bibfnamefont {S.~R.}\ \bibnamefont {White}},\
  and\ \bibinfo {author} {\bibfnamefont {E.~M.}\ \bibnamefont {Stoudenmire}},\
  }\bibfield  {title} {\bibinfo {title} {The \mbox{ITensor} software library
  for tensor network calculations},\ }\Eprint
  {https://arxiv.org/abs/2007.14822} {arXiv:2007.14822}  (\bibinfo {year}
  {2020})\BibitemShut {NoStop}%
\bibitem [{\citenamefont {Giardina}\ \emph {et~al.}(2010)\citenamefont
  {Giardina}, \citenamefont {Redig},\ and\ \citenamefont
  {Vafayi}}]{giardina2010correlation}%
  \BibitemOpen
  \bibfield  {author} {\bibinfo {author} {\bibfnamefont {C.}~\bibnamefont
  {Giardina}}, \bibinfo {author} {\bibfnamefont {F.}~\bibnamefont {Redig}},\
  and\ \bibinfo {author} {\bibfnamefont {K.}~\bibnamefont {Vafayi}},\
  }\bibfield  {title} {\bibinfo {title} {Correlation inequalities for
  interacting particle systems with duality},\ }\href
  {https://doi.org/10.1007/s10955-010-0055-0} {\bibfield  {journal} {\bibinfo
  {journal} {Journal of Statistical Physics}\ }\textbf {\bibinfo {volume}
  {141}},\ \bibinfo {pages} {242} (\bibinfo {year} {2010})}\BibitemShut
  {NoStop}%
\bibitem [{\citenamefont {Grosskinsky}\ \emph {et~al.}(2013)\citenamefont
  {Grosskinsky}, \citenamefont {Redig},\ and\ \citenamefont
  {Vafayi}}]{grosskinsky2013dynamics}%
  \BibitemOpen
  \bibfield  {author} {\bibinfo {author} {\bibfnamefont {S.}~\bibnamefont
  {Grosskinsky}}, \bibinfo {author} {\bibfnamefont {F.}~\bibnamefont {Redig}},\
  and\ \bibinfo {author} {\bibfnamefont {K.}~\bibnamefont {Vafayi}},\
  }\bibfield  {title} {\bibinfo {title} {Dynamics of condensation in the
  symmetric inclusion process},\ }\href {https://doi.org/10.1214/EJP.v18-2720}
  {\bibfield  {journal} {\bibinfo  {journal} {Electronic Journal of
  Probability}\ }\textbf {\bibinfo {volume} {18}},\ \bibinfo {pages} {1}
  (\bibinfo {year} {2013})}\BibitemShut {NoStop}%
\bibitem [{\citenamefont {Miller}(1974)}]{miller1974jackknife}%
  \BibitemOpen
  \bibfield  {author} {\bibinfo {author} {\bibfnamefont {R.~G.}\ \bibnamefont
  {Miller}},\ }\bibfield  {title} {\bibinfo {title} {The jackknife-a review},\
  }\href {https://doi.org/10.1093/biomet/61.1.1} {\bibfield  {journal}
  {\bibinfo  {journal} {Biometrika}\ }\textbf {\bibinfo {volume} {61}},\
  \bibinfo {pages} {1} (\bibinfo {year} {1974})}\BibitemShut {NoStop}%
\end{thebibliography}%

\end{document}